\documentclass[12pt]{article}
\usepackage{sqdec_new}
\usepackage{sqdec_figs}
\usepackage{cite}

\begin{document}
%Use symbols for the footnotes in the title page
\renewcommand{\thefootnote}{\fnsymbol{footnote}}

\thispagestyle{empty}

\begin{flushright}
{\parbox{4cm}{KA-TP-1-2002\\
UB-ECM-PF-02/03\\
MPI-PhT/2002-27\\
PSI-PR-02-07\\
hep-ph/0207364
}}
\end{flushright}

\vspace{0.5cm}

\begin{center}

{\Large \textbf{Fermionic decays of sfermions: a complete discussion at
    one-loop order}} 

\vspace{0.8cm}

{\large Jaume Guasch$^{a,b}$,  
Wolfgang Hollik$^{a,c}$, Joan Sol\`a$^{d,e}$}

\vspace*{0.8cm}

{\sl $^a$ Institut f\"ur Theoretische Physik, Universit\"at Karlsruhe,
  Kaiserstra{\ss}e 12, \\ D-76128 Karlsruhe, Germany\\
$^b$ Theory Group LTP, Paul
    Scherrer Institut, CH-5232 Villigen PSI, Switzerland\\
$^c$  Max-Planck-Institut f\"ur Physik,
   F\"ohringer Ring 6, D-80805 M\"unchen, Germany \\
$^d$  Departament d'Estructura i Constituents de la 
Mat\`eria,  Universitat de Barcelona, Diagonal 647, E-08028, Barcelona,
  Catalonia, Spain\\
$^e$ Institut de F\'{\i}sica d'Altes Energies, Universitat Aut\`onoma de
  Barcelona, E-08193, Bellaterra, Barcelona, Catalonia, Spain} 

\end{center}

\vspace*{0.5cm}

\begin{abstract} 
\noindent
We present a definition of an on-shell renormalization scheme for the
sfermion and chargino-neutralino sector of the Minimal Supersymmetric
Standard Model (MSSM). Then, apply this renormalization framework to the
interaction between charginos/neu\-tra\-li\-nos and sfermions. 
{A kind of universal corrections is identified, which allow to define
effective chargino/neutralino coupling matrices. In turn,  these
interactions generate (universal) non-decoupling terms that grow as the
logarithm of the heavy mass.} 
Therefore the full MSSM spectrum must be
taken into account in the computation of radiative corrections to
observables involving these interactions. {As an 
application we} analyze the full one-loop
electroweak radiative
corrections to the partial decay widths $\Gamma(\tilde{f}\to f\neut)$ and
$\Gamma(\tilde{f}\to f'\chi^\pm)$ for all sfermion flavours and
generations. These are combined with the QCD corrections to compute the
corrected branching ratios of sfermions. {It turns out that the}
electroweak corrections can
have an important impact on the partial decay widths, as well as the
branching ratios, in wide regions of the parameter space. The precise
value of the corrections is strongly dependent on the correlation
between the different particle masses.
\end{abstract}

% Set the footnote counter to zero
\setcounter{footnote}{0} 
%and use numbers
\renewcommand{\thefootnote}{\arabic{footnote}}
\newpage

\section{Introduction}

The Standard Model (SM) of the strong and electroweak interactions is the
present paradigm of particle physics. Its validity has been tested to a
level better than one per mille at particle
accelerators~\cite{SM2002}. 
Nevertheless, there are
arguments against the SM being the fundamental model of
particle interactions~\cite{Carena:2000yx}, giving rise to the
investigation of competing alternative or extended models, which
can be tested at high-energy colliders, such as the Large Hadron Collider
(LHC)~\cite{ATLASCMS}, or a $e^+e^-$ Linear
Collider (LC)~\cite{TESLATDR}. 
One of the most promising possibilities for physics beyond
the SM is the incorporation of Supersymmetry (SUSY), which leads to
a  renormalizable field theory with precisely calculable
predictions to be tested in future experiments. 
{The simplest supersymmetric extension of the SM is the Minimal Supersymmetric
Standard Model (MSSM)~\cite{MSSM}. Up to now the major effort on the
computation of SUSY radiative
corrections has been put into the computation of virtual SUSY effects in
observables that involve only SM external particles, or into the calculation of
loop effects in the extended
Higgs sector of the MSSM\footnote{See e.g.~\cite{IWQEMSSM}
and references therein.}. In this context, if the masses of the extra
non-standard particles
are very large as compared to the SM electroweak scale, the effects of
these particles decouple, leaving  the SM as a low-energy effective
theory\footnote{{See e.g.~\cite{Dobado:1997up} and references
therein.}}. This means that if the extra particles are
too heavy we could not discern between the SM and the MSSM by just
looking at the low-energy end of the spectrum, since the only trace of
the MSSM would be a light Higgs boson
($M_{h^0}\lsim 135\GeV$)~\cite{Higgs2L},
whose properties would not differ from the SM one.}
But for the case of direct production of SUSY particles, one also
needs a detailed knowledge of the higher-order effects for the processes
with these SUSY particles in the external states. 
{In contrast to the case {of  virtual} SUSY effects in SM Green's
functions, there is a great variety of additional electroweak MSSM
processes, viz. those involving Higgs bosons and/or (R-odd) sparticles
in the external legs, for which the decoupling limit cannot be
applied. In this case several kinds of non-decoupling effects may appear
which grow with the mass of the sparticles~\cite{DmbTeo,Hikasa:1996bw,Djouadi:1997wt,Katz:1998br}. These effects can be
very important, as they could provide the clue to discovering SUSY
physics in the colliders. This was amply demonstrated in the past for
decay processes~\cite{Coarasa:1996qa,Coarasa:1997ky}, 
and also very recently for production
cross-sections in hadron colliders\cite{Belyaev:2001qm}, 
in both cases exploiting
the SUSY threshold corrections in the top quark and Higgs boson
sector. In the present study, however, we will face not only threshold
effects, but also a new type of non-decoupling (so-called universal)
contributions. Both types of non-decoupling effects have to be
considered for a complete study of sfermion decays in the colliders.}
The LHC will be able to discover
new  particles with masses up to $2.5\TeV$.
Provided 
they are not too heavy, the LC will be able to
make precision measurements of their properties.
For example, at a $500\GeV$ {LC with} a total integrated luminosity of
$500\ {\rm fb}^{-1}$ a measurement of the top-squark mass and the
top-squark mixing angle can be performed with a precision
of $0.5\%$ and $1.5\%$ respectively~\cite{Berggren:1999ss}.
For an adequate analysis, precise theoretical predictions are
required, going beyond the Born {approximation.
These studies of purely
  supersymmetric processes at the quantum level are necessary not only
  to refine the prediction of the corresponding observables but also
  because  1) the quantum corrections may severely affect the physical
  production of the supersymmetric particles, and 2) in addition they
  probe the underlying SUSY nature present in the model, that is: the
  relation between the gauge couplings of the SM gauge bosons and the
  Yukawa coupling of its SUSY partners (charginos and
  neutralinos). Beyond leading order this relation receives corrections
  which are non-decoupling.}

A number of studies have already addressed this issue, 
for production as well as for decay processes.
For squark and gluino production in hadron collisions,
the  NLO QCD corrections 
are available~\cite{Beenakker:1997ch}; 
for squark-pair production
in $e^+e^-$ collisions, the NLO QCD 
are also known, together with the Yukawa corrections~\cite{Eberl:1996wa}. 
Concerning the subsequent
squark decays into charginos/neu\-tra\-li\-nos, the QCD corrections were
presented in~\cite{Kraml:1996kz,Djouadi:1997wt}\footnote{The gluino
decay  channel, which can be overwhelming for $m_{\tilde
  q}>m_q+m_{\tilde g}$, was studied in~\cite{Beenakker:1996de}. Here we
will assume $m_{\tilde g}>m_{\tilde q_a}$.},
whereas the Yukawa corrections were given
in~\cite{Guasch:1998as}\footnote{A direct comparison between the QCD and
the Yukawa corrections can be found in Ref.\cite{Guasch:2000up}.}.
In this last work large corrections were found. 
They were derived, however, 
in the \textit{higgsino} approximation for the chargino; 
hence, a full computation is required 
to consolidate the significance of the loop effects.

We have performed a complete one-loop computation of the electroweak
radiative corrections to the partial decay widths of 
sfermions  
into fermions and charginos/neutralinos, 
\begin{equation}
\Gamma(\tilde{f} \to f' \chi)\,\,.
\label{eq:gammadef}
\end{equation}
We present the structure of the corrections in detail, 
and illustrate their main features and their significance 
in representative numerical examples. 
Explicit results 
are displayed for all kind of sfermions. 
First results of this study were presented in Ref.\cite{Guasch:2001kz}.

In processes with exclusively SM particles in external states,
it is possible to divide the  one-loop contributions into
SM-like and non-SM-like subclasses.
This separate treatment is often used in the literature, 
and it is useful since it allows to make the
computation in small steps, checking each sector individually.
As a distinctive feature of the radiative corrections to processes with
supersymmetric particles in the external legs, this separability is lost.
In such kind of processes the ultraviolet (UV)
divergences of diagrams with virtual SM particles cancel the
UV divergences of diagrams with non-SM particles. 
Any partial computation would yield
UV-divergent and thus meaningless results.
{For this reason we have to compute the entire set of one-loop
   contributions for our processes}, with the proper counterterms involving the
  renormalization of almost the full MSSM Lagrangian.
As a direct consequence,
many non-decoupling effects appear.  {Moreover, we remark that
  since we have sparticles in the external legs a consistent calculation
  of the loop integrals requires the use of dimensional reduction in
  order that the {regularization} 
procedure preserves supersymmetry~{\cite{Dred}}\footnote{{One could also make
    use of a SUSY-breaking regularization, and introduce
    corresponding SUSY-breaking counterterms to restore supersymmetry at
    the one-loop order, see e.g.~\cite{HollikDominik}.}}.}

Section~\ref{sec:renor} contains  the renormalization of the
sfermion sector (section~\ref{sec:renorsf}), the chargino/neutralino sector
(section~\ref{sec:renorcn}), and their interaction
(section~\ref{sec:renorlagrangian}). The numerical analysis in done in
section~\ref{sec:numeric}, including the combination of the electroweak
effects with the QCD ones, and the computation of the one-loop corrected
branching ratios. Finally section~\ref{sec:conclu} is devoted to the
conclusions.

\section{Renormalization and radiative corrections}
\label{sec:renor}
\subsection{Introduction}

It is our aim to complement the on-shell scheme of the SM to include the
SUSY particles. The renormalization of the SM is done 
{according to Ref.~\cite{Bohm:1986rj} apart from some sign
 conventions. However, when 
 extending this renormalization framework to embrace the whole MSSM we
 will treat the field renormalization of the supersymmetric particles in
 a different way that will be described at due point.} 
The so-called $\alpha$-scheme is used, in which
the input parameters for the gauge sector are chosen to be the fine
structure constant $\alpha$ (defined in the Thomson limit) {and}
 the pole
masses of the weak gauge bosons $\mw$, $\mz$. The electroweak mixing angle
is defined 
on-shell:\footnote{We abbreviate trigonometric functions by their
  initials, like 
  $\sw\equiv\sin\theta_W$,
  $\stbt\equiv\sin(2 \beta)$, $t_W=\sw/\cw$, etc. }
$\sws=1-\mws/\mzs$. 

The Higgs sector of the MSSM has received a lot of
attention in the literature~\cite{Higgs2L,Dabelstein:1995hb,andtheothers}. Here
we follow Ref.\cite{Dabelstein:1995hb}. In fact, the only 
ingredient of the Higgs sector needed  at one-loop order for
renormalization of 
the sfermion and the chargino-neutralino sectors is the renormalization
of $\tb$. The counterterm is determined by the condition for the
counterterms of the two vacuum expectation values,
$$
\frac{\delta v_1}{v_1}=\frac{\delta v_2}{v_2}\,\,,
$$
together with the absence of mixing between the on-shell $A^0$ Higgs boson and
the $Z$ weak gauge boson, which gives
$$
\frac{\delta \tb}{\tb}=-\frac{1}{\mz\stbt} \Sigma^{A^0Z^0}(\mAs)\,\,.
$$
{Our choice of $\tan\beta$ is based on simplicity. It is known that all
definitions of $\tan\beta$ not directly related to a physical observable
are subject to some gauge dependence and/or induce large variations of
the corrections with the parameters\footnote{{For a specific physical
definition of $\tan\beta$, free of these problems, see
e.g. Ref.~\cite{Coarasa:1996qa}. As 
a drawback, however, one has to compute the process-dependent
corrections. See also~\cite{Freitas:2002um} for a recent review on this
subject.}}. In our case the dependence is small and, with this simple 
and well tested definition, we can avoid introducing process-dependent
corrections in a framework, like ours, which is already quite
cumbersome. Needless to say, the physical observables we are addressing,
like decay rates and branching ratios, are completely insensitive to our
particular choice of $\tan\beta$. }

Besides parameter renormalization,
we introduce field-renormalization constants for each particle. 
Exploiting the freedom in the treatment of field renormalization,
they have been chosen in a way to get fairly simple expressions for the 
physical observables under study.

\subsection{Sfermion sector}
\label{sec:renorsf}

{Throughout} this paper we will be using the 3th family squarks as a generic
fermion-sfermion notation. The same relations hold for sleptons, by
changing {the corresponding charges appropriately}.

We denote {the two sfermion-mass eigenvalues
by $m_{\tilde{f}_a}\,(a=1,2)$, with
$m_{\tilde{f}_1}>m_{\tilde{f}_2}$}.
The sfermion-mixing angle $\osf$
is defined by the transformation relating the weak-interaction
($\sfr^\prime_a=\sfr_L, \sfr_R$) and the mass eigenstate
 ($\sfr_a=\sfr_1, \sfr_2$) sfermion bases:
\begin{equation}
\label{eq:defsq}
  \sfr^\prime_a=R_{ab}^{(f)}\, \sfr_b\,\,; \,\,\,\,
  R^{(f)}=\left(\begin{array}{cc}
      \cos\osf&-\sin\osf \\
      \sin\osf &\cos\osf
    \end{array}\right)\,.
\end{equation}
By this  basis transformation, the sfermion mass matrix,
\begin{equation}
{\cal M}_{\tilde{f}}^2 =\left(\begin{array}{cc}
M_{\tilde{f}_L}^2+m_f^2+\ctbt(T_3-Q\,s_W^2)\,M_Z^2 
 &  m_f\, M^{LR}_f\\
 m_f\, M^{LR}_f &
 M_{\tilde{f}_R}^2+m_f^2+Q\,\ctbt \,s_W^2\,M_Z^2  
\end{array} \right)\,,
\label{eq:sbottommatrix}
\end{equation}
becomes diagonal: 
$R^{(f)\dagger}\,{\cal M}_{\tilde{f}}^2\,R^{(f)}=
{\rm diag}\left\{m_{\tilde{f}_1}^2,
  m_{\tilde{f}_2}^2\right\}$. $M_{\tilde{f}_L}^2$ is the
soft-SUSY-breaking mass parameter of the $SU(2)_L$
doublet\footnote{{With $M_{\tilde{t}_L}=M_{\tilde{b}_L}$ due to $SU(2)_L$ gauge invariance.}}, whereas
$M_{\tilde{f}_R}^2$ is the soft-SUSY-breaking mass parameter of the
singlet. $T_3$ and 
$Q$ are the usual third component of the isospin and the
{electric} 
charge respectively. The mixing {parameters} in the non-diagonal entries {read}
$$
M^{LR}_b=A_b-\mu\tb\ \ \ ,\ \ \ M^{LR}_t=A_t-\mu/\tb\,\,.
$$

Our aim is to compute the radiative corrections in an on-shell
renormalization scheme; hence, the input parameters are  physical observables (i.e.\ the physical masses
$m_{\tilde{b}_2}, m_{\tilde{b}_1}$, \ldots) rather than formal parameters in the
 Lagrangian
(i.e.\ the soft-SUSY-breaking parameters $M_{\tilde{b}_L}^2, A_b$, \ldots\ in
eq.~(\ref{eq:sbottommatrix})). Specifically, we  use the following set of
independent parameters for the squark sector:
\begin{equation}
(m_{\tilde{b}_1}, m_{\tilde{b}_2}, \osb ,m_{\tilde{t}_2}, \ost)\,.
\label{eq:inputb}\label{eq:inputt}
\end{equation}
The value of the other
stop mass, $m_{\tilde{t}_1}$, {is derived
from this set of input parameters.}
The sbottom and stop trilinear soft-SUSY-breaking terms $A_b$ and $A_t$
are fixed at the 
tree-level by the previous parameters as follows:
\begin{equation}
A_{b}=\mu\,\tan\beta+
{m_{\tilde{b}_1}^2-m_{\tilde{b}_2}^2\over 2\,m_b}\,\sin{2\,\osb}\,
;
\ \ \ \
A_{t}=\mu\,\cot\beta+
{m_{\tilde{t}_1}^2-m_{\tilde{t}_2}^2\over 2\,m_t}\,\sin{2\,\ost}\,
.
\label{eq:Abt}
\end{equation}
We impose the approximate (necessary) condition
\begin{equation}
A_q^2<3\,(m_{\tilde{t}}^2+m_{\tilde{b}}^2+M_H^2+\mu^2)\,,
\label{eq:necessary}
\end{equation}
where $m_{\tilde{q}}$ is of the order of the average squark masses
for $\tilde{q}=\tilde{t},\tilde{b}$, to avoid colour-breaking minima 
in the MSSM Higgs potential\,\cite{Frere:1983ag}. Of course the relation~(\ref{eq:Abt})
receives one-loop corrections; however, since these parameters do not enter the
tree-level expressions, these effects translate into two-loop corrections to the
process under study. The bound~(\ref{eq:necessary}) translates into a stringent
constraint to the sbottom-quark mixing angle for moderate and large values of
$\tb \gsim 10$: with an approximate limit $|\mu|\gsim 80\GeV$  from the
negative output of the chargino search at LEP,  the
condition~(\ref{eq:necessary}) can only be satisfied by a cancellation of the
two terms in~(\ref{eq:Abt}) which is easily spoiled when $\osb$ is
varied. {However, the} right hand side of
eq.~(\ref{eq:necessary}) is not rigorous~{\cite{Frere:1983ag}};
so we 
will present results also when this bound is not satisfied, but we will clearly
mark these regions. With the use of the bound~(\ref{eq:necessary}) also the
squark-squark-Higgs-boson couplings are restricted. This is a welcome feature,
since these couplings can 
in general be very large, eventually spoiling perturbativity.

The slepton sector follows the same procedure, but only one sneutrino is
present. Thus the input parameters in eq.~(\ref{eq:inputb}) are reduced
to the charged slepton masses and mixing angle.

The definition of the on-shell renormalization scheme is driven by the
input parameters~(\ref{eq:inputb}). The two bottom-squarks and the
lightest top-squark are defined to be on-shell, whereas the heaviest
top-squark mass receives quantum corrections. The mixing angle
renormalization must also be given. Since there is 
no unique concept
of an
\textit{on-shell angle}, a practical definition is given, which is
general enough to be used in any sfermion observable.

Other definitions of an
\textit{on-shell} scheme are also conceivable.
One could
think, for example, of a concept having all the sfermions defined
on-shell, so that the input parameters would be the four masses and one
mixing angle. 
Such a scheme, however, is problematic. First of all, this
input parameter set is not complete, and one needs to give also the sign
of the angle to determine completely the parameters of the sfermion
section. Second, 
the expression derived for the counterterm of the mixing angle is
  not well defined when the angle is zero (no mixing). Admittedly, the
  zero-mixing-angle case could be thought of as an academic limit;
  it is, however, a very useful scenario for the study of sfermions at
  colliders, and has widely been used accordingly. 
  Therefore, a renormalization
  framework which permits a consistent treatment of the 
  zero-mixing-angle limit is desirable.

This on-shell sfermion-sector renormalization was already introduced
in~\cite{Guasch:1998as}. Here we expand this renormalization framework
to include the relation between bottom-squark and top-squark
counterterms, and 
{provide a thorough discussion.}

For each squark {type $a=1,2$} we introduce a set of
{field-renormalization} constants as follows,
\begin{equation}
  \sfr_a^{(0)} = (Z^a_{\tilde f})^{1/2}\,\sfr_a + 
 (1-\delta^{ab})\,\delta Z^{ab}_{\tilde f} \sfr_b\ \ \ \,,
\label{eq:sqwavecounter}
\end{equation}
where we have attached a superscript $^{(0)}$ to the bare fields, {and
$\delta^{ab}$ is the usual Kronecker delta}. As for 
parameter renormalization, we introduce counterterms for each
independent parameter in eq.~(\ref{eq:inputb}):
\begin{equation}
  \label{eq:sqcounter}
  (m_{\sbottom_a}^2)^{(0)}\equiv m_{\sbottom_a}^2+\delta m_{\sbottom_a}^2
\,\,,\,\,  (m_{\stopp_2}^2)^{(0)}\equiv m_{\stopp_2}^2+\delta m_{\stopp_2}^2
\,\,,\,\,  \osf^{(0)}\equiv \osf+\delta \osf \,\,.
\end{equation}

The bare fields in the interaction basis are related to the bare fields
in the mass basis as
\begin{equation}
  \label{eq:bareweakmass}
  (\sfr')^{(0)}=R^{(f)(0)} (\sfr)^{(0)} \,\,,\,\,R^{(f)(0)}\equiv R^{(f)}+\delta R^{(f)}\,\,,\,\,\delta
  R_{ab}^{(f)}=\frac{\partial R_{ab}^{(f)}}{\partial \osf} \delta \osf\,\,.
\end{equation}
The counterterms to the soft-SUSY-breaking parameters in
eq.~(\ref{eq:sbottommatrix}) can be found by
\begin{equation}
  \label{eq:sqdmweak}
  ({\cal M}_{\tilde f}^2)^{(0)}= R^{(f)(0)} 
  \left(\begin{array}{cc}
      (m_{\tilde f_1}^2)^{(0)} & 0 \\
      0 &      (m_{\tilde f_2}^2)^{(0)}  
    \end{array}
  \right)(R^{(f)(0)})^\dagger\,\,.
\end{equation}
The bottom-squarks and the lightest top-squark are defined to be
on-shell, the residue of the {renormalized} propagators is taken
to be $1$ and we 
require no-mixing between the sfermions, that is\footnote{{It is
    understood that only the real part of the self-energies is taken in
    the counterterm definitions.}}
\begin{eqnarray}
  \delta m^2_{\tilde f}&=&-\Sigma_{\tilde f}(m^2_{\tilde  f})
  \ \ , \ \ \tilde{f}=\sbottom_a,\stopp_2\,\,,\nonumber\\
  \delta Z^a_{\tilde f}&=& \Sigma'_{\tilde f_a}(m^2_{\tilde f_a})  \ \ , \ \ \tilde{f}_a=\sbottom_a,\stopp_a\,\,,\nonumber\\
   \delta Z^{ab}_{\tilde f}&=&\frac{\Sigma_{\tilde
  f}^{ab}(\msfbs)}{\msfbs-\msfas}\ \ , \ \ \tilde{f}=\tilde{b},
  \tilde{t}\ \ ,\ \   (a\neq b)\,\,. 
 \label{eq:onshellsfermions}
\end{eqnarray}
{The mixing angles in the sfermion sector do not receive radiative
corrections. The corresponding counterterm is fixed by means of the
{non-diagonal field-renormalization constants of} eq.~(\ref{eq:sqwavecounter}) as follows: }
\begin{equation}
\delta\osf=\frac{1}{2}\,(\delta Z^{12}_{\tilde{f}}-\delta Z^{21}_{\tilde{f}}) 
=\frac{1}{2}\,\frac{\Sigma^{12}_{\tilde{f}}(\msfts)+\Sigma^{12}_{\tilde {f}}(\msfos)}{\msfts-\msfos} \,. 
\label{eq:deltatheta}
\end{equation}
The set of equations~(\ref{eq:onshellsfermions}), (\ref{eq:deltatheta})
defines all the counterterms in the sfermion sector. 
{The one-loop electroweak Feynman diagrams contributing to the
  sfermions self-energies are shown schematically in
  Fig.~\ref{diags:self}a.} 

The heaviest top-squark is not defined on-shell, and therefore its mass
receives radiative corrections. In order to find them we make use of
eq.~(\ref{eq:sqdmweak}). 
Since the bottom-squarks are defined to be on-shell, the counterterm to
the soft-SUSY-breaking squared mass $M^2_{\tilde b_L}$ is found to be
\begin{equation}
  \label{eq:deltamsqLL}
  \delta M^2_{\tilde{b}_L}=\delta {(\cal M}_{\tilde b}^2)_{11} - 2 \mb
  \delta\mb - \ctbt (T_3^b -Q^b s_W^2 ) \delta \mzs -
  \mzs \left( \delta \ctbt (T_3^b -Q^b s_W^2 ) -  \ctbt
  Q^b\delta s_W^2 \right)\,\,,
\end{equation}
which is used in the top-squark matrix
\begin{equation}
  \label{eq:deltamsq11top}
  \delta {(\cal M}_{\tilde t}^2)_{11}= \delta M^2_{\tilde{b}_L} + 2 \mt
  \delta\mt + \ctbt (T_3^t -Q^t s_W^2 ) \delta \mzs +
  \mzs \left( \delta \ctbt (T_3^t -Q^t s_W^2 ) -  \ctbt
  Q^t\delta s_W^2 \right)\,\,.
\end{equation}
{The fermion mass-counterterms $\delta \mb,\delta \mt$ are
  computed with the help of the Feynman diagrams of Fig.~\ref{diags:self}b.}
With {these} results, the counterterm for the $\stopp_1$ mass is
  found to be
\begin{equation}
  \label{eq:dmstop1}
  \delta m_{\stopp_1}^2=\frac{1}{(R_{11}^{(t)})^2} \left(
    \delta {(\cal M}_{\tilde t}^2)_{11} - (R_{21}^{(t)})^2 \delta
    m_{\stopp_2}^2 -2 \msto^2 R_{11}^{(t)} \delta R_{11}^{(t)} - 2
    \mstt^2 R_{12}^{(t)}
    \delta R_{12}^{(t)}
    \right)\,\,.
\end{equation}
Note that this renormalization prescription
breaks down for $\ost=\pi/2$. In this case we have that $\stopp_1=\stopp_R$,
whose mass is not related to the bottom-squark; then it would be better
to use a renormalization prescription where $\msto$ is the input
parameter. Of course, this renormalization prescription would break down
in turn at $\ost=0$. 

The one-loop on-shell mass for the $\stopp_1$ is then given by
\begin{equation}
  \label{eq:stop1loopmass}
  (\msto^2)^{\rm os}= (\msto^2)^{tree}+\delta
  \msto^2+\Sigma_{\stopp_1} (\msto^2)\,\,.
\end{equation}
In Section~\ref{sec:numeric} we give numerical values for the correction
to the top-squark mass. 

\figone

\subsection{Chargino-neutralino sector}
\label{sec:renorcn}
This sector contains six particle masses, but only 
three free parameters are available for an independent renormalization.
As a consequence, we are not allowed  to impose
on-shell conditions for all the particle masses.
For the independent input parameters, we choose:
the masses of the two charginos and the mass of the lightest neutralino,
\begin{equation}
(M_1, M_2, M_1^0)\,.
\label{eq:inputINOS}
\end{equation}

Although the tree-level chargino-neutralino sector is well known, we
give here a short description, in order to set our conventions. 
We start by constructing the following set of Weyl spinors:
\begin{equation}
\begin{array}{lcl}
\Gamma^+&\equiv&(-i \tilde W^+,\tilde H_2^+) \,\,,\\
\Gamma^-&\equiv&(-i \tilde W^-,\tilde H_1^-) \,\,,\\
\Gamma^0&\equiv&(-i \tilde B^0,-i \tilde W_3^0,\tilde H_1^0,\tilde H_2^0) \,\,.
\end{array}
\label{eq:inosweak}
\end{equation}
The mass Lagrangian in this basis reads
\begin{equation}
{\cal L}_M=-\frac{1}{2}\pmatrix{\Gamma^+,\Gamma^-}
\pmatrix{0&{\cal M}^T\cr
{\cal M}&0}
\pmatrix{ \Gamma^+\cr \Gamma^-}
-\frac{1}{2} \pmatrix{\Gamma_1,\Gamma_2,\Gamma_3,\Gamma_4}
{\cal M}^0 \pmatrix{\Gamma_1\cr \Gamma_2 \cr \Gamma_3 \cr \Gamma_4}
+\mbox{ h.c.}\,\,,
\end{equation}
where we have defined
\begin{eqnarray}
{\cal M}&=&\pmatrix{
M&\sqrt{2} M_W \sbtt \cr
\sqrt{2} M_W \cbtt&\mu
}\,\,,\nonumber\\
{\cal M}^0&=&
\pmatrix{
M^\prime&0&M_Z\cbtt\swp&-M_Z\sbtt\swp \cr
0&M&-M_Z\cbtt\cwp&M_Z\sbtt\cwp\cr
M_Z\cbtt\swp&-M_Z\cbtt\cwp&0&-\mu \cr
-M_Z\sbtt\swp&M_Z\sbtt\cwp&-\mu&0
}\,\,\,,
\label{eq:mmassacplus}\label{eq:mmassaneut}\end{eqnarray}
{with $M$ and $M'$} the $SU(2)_L$ and $U(1)_Y$
soft-SUSY-breaking gaugino 
masses. 
The four-component mass-eigenstate fields are related to the ones
in~(\ref{eq:inosweak}) by 
\begin{equation}
  \chi_i^{+}= \pmatrix{
    V_{ij}\Gamma_j^{+} \cr U_{ij}^{*}\bar{\Gamma}_j^{-}
    }
  \; \;\; \;\;,\;\;\;\;\;
  \chi_i^{-}= {\cal C}\bar{\chi_i}^{+T} =\pmatrix{
    U_{ij}\Gamma^{-}_j \cr V_{ij}^{*}\bar{\Gamma}_j^{+} 
    }% \ ,  
  \label{eq:cinos}
%   \nonumber
\ \ , \ \  \chi_{\alpha}^0= \pmatrix{N_{\alpha\beta}\Gamma_{\beta}^0 \cr 
    N_{\alpha\beta}^{*}\bar{\Gamma}_{\beta}^0} =  
  {\cal C}\bar{\chi}_{\alpha}^{0T}\ ,
  \label{eq:ninos} 
  \nonumber
\end{equation}
where $U$, $V$ and $N$ are in general complex matrices that 
diagonalize the mass-matrices~(\ref{eq:mmassacplus}):
\begin{equation}
\begin{array}{lcccl}
U^* {\cal M} V^\dagger&=&{\cal M}_D&=&{\rm diag}\left(M_1,M_2\right)\,\,(0<M_1<M_2)\,\,,\\
 N^*{\cal M}^0 N^\dagger &=&{\cal M}^0_D&=&
{\rm diag}\left(M_1^0,M_2^0,M_3^0,M_4^0\right)\,\,(0<M_1^0<M_2^0<M_3^0<M_4^0)\,\,.
\end{array}\label{eq:defUVN}
\end{equation}

The renormalization framework is as follows: we introduce
the renormalization constants for each parameter in the mass
matrices~(\ref{eq:mmassacplus}). The counterterms $\delta \mz$,
$\delta\mw$, $\delta\cw$ and $\delta\tb$ are fixed from the conditions
on the gauge and Higgs sectors of the theory, and we must provide
renormalization prescriptions for $\delta M$, $\delta \mu$ and 
$\delta M'$. The mixing matrices $U$, $V$ and $N$ are defined to
diagonalize the renormalized mass matrices. The bare mass Lagrangian is
then
\begin{equation}
  \label{eq:baremasscn}
  {\cal L}^{(0)}_M=- (\bar{\chi}^-)^{(0)} \left( M_D  + 
  \delta M_D  \right)(\chi^-)^{(0)}
  -\frac{1}{2}
  (\bar{\chi}^0)^{(0)} \left( M_D^0 +
  \delta M_D^0 \right)  (\chi^0)^{(0)}
\end{equation}
with 
$$
\delta M_D=U^* \delta{\cal M} V^\dagger \ \ , \ \ \delta M_D^0 = N^*
\delta {\cal M}^0 N^{\dagger}\,\,.
$$
Note that these counterterm matrices are non-diagonal. This
renormalization corresponds to the following relation between the bare
weak- and mass-eigenstates fields:
\begin{equation}
  (\chi_i^{-})^{(0)}= 
  \pmatrix{U_{ij}(\Gamma_j^{-})^{(0)} \cr V_{ij}^{*}(\bar{\Gamma}_j^{+})^{(0)}}
  \ \ , \ \  (\chi_{\alpha}^0)^{(0)}= 
  \pmatrix{N_{\alpha\beta}(\Gamma_{\beta}^0)^{(0)} \cr 
    N_{\alpha\beta}^{*}(\bar{\Gamma}_{\beta}^0)^{(0)}} \ .
  \label{eq:ninosbare} 
  \nonumber
\end{equation}
Note that no counterterms for the mixing matrices are introduced.
As far as field renormalization is concerned, 
we introduce different field-renormalization constants for each chargino
and neutralino:
\begin{equation}
  \label{eq:fieldcharneut}
  (\chi_s)^{(0)}\equiv\chi_s+\frac{1}{2} \delta Z_L^s \pl \chi_s + \frac{1}{2}
  \delta Z_R^s \pr \chi_s\ \ ,\ \  \chi_s\equiv \chi^{\pm}_i,
  \chi^0_\alpha\,\,, 
\end{equation}
{with the chirality projectors $P_{\{L,R\}}=1\mp\gamma_5$.}
The renormalization framework is thus complete\footnote{{One can
    take either the $\chi^+$ or the $\chi^-$ to perform the
    renormalization. We will take 
    the self-energies and the corresponding renormalization constants to
    be that of the $\chi^-$. One has, evidently, $\delta Z_L^{-i}=\delta
    Z_R^{+i}$ and  $\delta Z_L^{+i}=\delta Z_R^{-i}$.}}. Now we must
    {supply 
renormalization} conditions for each parameter {in
  eq.~(\ref{eq:inputINOS})} and {wave-function}
renormalization constant.
We will {fix 
 $\delta M$ and $\delta \mu$ by} requiring that the one-loop
renormalized chargino masses are the on-shell masses. $\delta M'$ is fixed by
requiring that the lightest renormalized neutralino mass is the
on-shell mass. The {wave-function} renormalization constants are fixed by 
requiring that the renormalized propagator has residue one.

Using this framework, we are able to relate the counterterms of the
fundamental parameters to the mass-counterterms $\delta M_i$ of the
charginos. Similarly to~\cite{Pierce:1994gj} 
we find:
\begin{eqnarray}
M_1 \,\delta M_1 + M_2 \,\delta M_2 &=&  M \,\delta M + \mu \,\delta\mu+ \delta
\mws , \nonumber\\
M_1 M_2
\left(M_1 \,\delta M_2+M_2\,\delta M_1\right)&=&\left(M \mu -\mws
  \stbt\right)\bigg[M \,\delta \mu + \mu \,\delta M\nonumber\\
&&-\mws \,\delta\stbt-\stbt \,\delta\mws\bigg] .
  \label{eq:defcountcha} 
\end{eqnarray}
The mass counterterms
$\delta M_i$ are fixed using the on-shell scheme
relation, in the convention 
of~\cite{Bohm:1986rj} (but with opposite sign for $\Sigma$), 
\begin{equation}
\frac{  \delta M_i}{M_i}=-\frac{1}{2}\left(\Sigma_L^{-i}(M_i^2)+\Sigma_R^{-i}(M_i^2)\right)-\Sigma_S^{-i}(M_i^2)\,\,,
  \label{eq:defdeltami}  
\end{equation}
where $\Sigma_{\{L,R,S\}}^i$ denote
the one-loop unrenormalized left-, right-handed and scalar components of
the self-energy for the $i$th-chargino. The wave function
renormalization constants are
\begin{eqnarray} 
  \delta Z_{L,R}^{-i} &=& \Sigma^{-i}_{L,R}(M_i^2)+M_i^2\,[\Sigma^{-i\,\prime}_L 
  (M_i^2)+\Sigma^{-i\,\prime}_R(M_i^2) 
  +2\,\Sigma^{-i\,\prime}_S(M_i^2)]\,.
  \label{eq:DSRC} 
\end{eqnarray} 
{The one-loop diagrams contributing to the chargino/neutralino
  self-energies are shown schematically in Fig.~\ref{diags:self}c.}

By solving the equations~(\ref{eq:defcountcha}) we find the counterterms
of the independent mass parameters of the chargino mass matrix~(\ref{eq:mmassacplus}):
\begin{eqnarray}
           \delta M&=&\frac{M \,X_1-\mu \,X_2}{M^2-\mu^2}\ \ , \ \ 
         \delta \mu=\frac{\mu\, X_1-M\, X_2}{\mu^2-M^2}\ \ , \nonumber\\
         X_1&=&M_1 \,\delta M_1+M_2 \,\delta M_2-\delta \mws\ \ , \ \ \nonumber\\
         X_2&=&\frac{M\, \mu -\mws \, \stbt}{M_1\, M_2}\left(M_1 \,\delta M_2+M_2\,\delta M_1\right)
          +\mws\,\delta \stbt+\stbt\,\delta\mws\,\,.
  \label{eq:soldM2dmu}
\end{eqnarray}
$\delta M'$ is determined from  the lightest neutralino
mass, inverting the relation
\begin{equation}
  \label{eq:defcountneut}
N^*_{1 \alpha} \delta{\cal M}^0_{\alpha\beta} N_{ 1\beta}^{*} = \delta M_1^0\,\,,
\end{equation}
where the neutralino-mass counterterm 
$\delta M_1^0$ is fixed by the on-shell condition for $\chi^0_1$,
in analogy to~(\ref{eq:defdeltami}). The result is then
\begin{equation}
  \label{eq:soldM1}
  \delta M'=\frac{1}{N_{11}^{*2}}\left(\delta M_1^0 -
  \sum_{\alpha\ {\rm or}\ \beta\neq1}N^*_{1 \alpha} \delta{\cal M}^0_{\alpha\beta} N_{
  1\beta}^{*} 
  \right)\,\,.
\end{equation}

It is a
non-trivial check that with the counterterms determined  in
eqs.~(\ref{eq:defcountcha}) and (\ref{eq:defcountneut}), the 
one-loop masses for the {remaining} neutralinos,
computed as the pole masses, are UV-finite.
The one-loop on-shell neutralino masses read
\begin{equation}
  M_\alpha^{0 \ {\rm os}}=M_\alpha^0+N^*_{\alpha\beta} \delta{\cal M}^0_{\beta\gamma} N_{\alpha\gamma}^{*}
  +M_\alpha^0\,\left\{\frac{1}{2}\left(\Sigma_L^\alpha(M_\alpha^{02})+\Sigma_R^\alpha(M_\alpha^{02})\right)+\Sigma_S^\alpha(M_\alpha^{02})\right\}\,\, ,
  \label{eq:defmass1l}  
\end{equation}
where now the parameters of eq.~(\ref{eq:mmassaneut}) and the masses and
mixing matrices computed in~(\ref{eq:defUVN}) have to be regarded as
\textit{renormalized} quantities.

The choice of the lightest neutralino to fix the counterterm 
$\delta M'$ in~(\ref{eq:defcountneut}) is only efficient if it has
a substantial \textit{bino} component. If $M'\gg(|\mu|,M)$ then
$|N_{11}|\ll1$, and the extraction of $\delta M'$
from~(\ref{eq:soldM1}) would amplify the radiative corrections
artificially. In this case
it would be better to extract $\delta M'$ from the
$\alpha$th neutralino, such that $|N_{1\alpha}|$ is large. 
This is, however, not relevant for the scenarios which are discussed
in this work.
Notice also that our renormalization procedure 
makes use of positive-definite mass eigenvalues for charginos and
neutralinos, which require the introduction of some purely-imaginary 
non-zero elements in the $N$-matrix~(\ref{eq:defUVN}). Had we
chosen a 
real $N$-matrix, with some negative eigenvalues, the various
renormalization conditions would be plagued with the explicit sign of the
corresponding eigenvalue (see e.g.~\cite{Guasch:1997dk}). 

At one-loop, also
mixing self-energies between the different neutralinos and charginos are
generated, which we write as follows:
\begin{equation}
  -i \hat\Sigma^{\alpha\beta}(k^2)=-i \left(\hat \Sigma^{\alpha\beta}_L(k^2)
   \slas{k} \PL + \hat \Sigma^{\alpha\beta}_R(k^2)   \slas{k} \PR
   +\hat\Sigma^{\alpha\beta}_{SL}(k^2) \PL + \hat\Sigma^{\alpha\beta}_{SR}(k^2) \PR\right)\,\,,\,\,\alpha\neq\beta\,\,,
  \label{eq:mixingneut}
\end{equation}
with
$\hat\Sigma$ denoting the renormalized two-point functions. 
For the neutralinos,
the renormalized self-energies~(\ref{eq:mixingneut}) are related to the
unrenormalized ones according to
\begin{equation}
  \hat\Sigma^{\alpha\beta}_{\{L,R\}}=\Sigma^{\alpha\beta}_{\{L,R\}}\,\,,\,\,
  \hat\Sigma_{SL}^{\alpha\beta}=\Sigma_{SL}^{\alpha\beta}{
    -N_{\alpha\gamma}\delta{\cal M}^{0*}_{\gamma\lambda} N_{\beta\lambda}}\,\,,\,\,
  \hat\Sigma_{SR}^{\alpha\beta}=\Sigma_{SR}^{\alpha\beta}{
    -N^*_{\alpha\gamma} \delta{\cal M}^{0}_{\gamma\lambda} N_{\beta\lambda}^{*}}\,\,.
  \label{eq:mixinneutren}  
\end{equation}
{Analogous expressions hold for the $\cmin$ charginos, replacing
$(\alpha\beta)\to(ij)$ in eq.~(\ref{eq:mixingneut}), the renormalized
$\cmin$ chargino self-energies being given by}
\begin{equation}
  \label{eq:mixingchar}
    \hat\Sigma^{-ij}_{\{L,R\}}=\Sigma^{-ij}_{\{L,R\}}\,\,,\,\,
  \hat\Sigma_{SL}^{-ij}=\Sigma_{SL}^{-ij}{
    -U_{ik}^*\delta{\cal M}_{kl} V_{jl}^*}\,\,,\,\,
  \hat\Sigma_{SR}^{-ij}=\Sigma_{SR}^{-ij}{
    -V_{ik} \delta{\cal M}_{lk} U_{lj}}\,\,.
\end{equation}

The one-loop mixing {self-energies 
also contribute} to the chargino
and neutralino masses;  their contribution is, however, of higher order in
perturbation theory, and we do not take it into account in the mass spectrum.

The contribution of these mixing self-energies to the one-loop decay
form factors can be written as follows. If $\tilde T_\alpha$ is the
amputated one-particle irreducible 3-point Green's function for the
creation of a 
$\neut_\alpha$ (represented by a spinor $v_\alpha$), the
full one-loop process amplitude reads: 
\begin{equation}
  \label{eq:externalZ}
  T_\alpha=\bar{u} \tilde T_\alpha v_\alpha+\sum_{\beta\neq\alpha}\bar{u} \tilde T_\beta ({\cal Z}_L^{0\beta\alpha}\pl + {\cal Z}_R^{0\beta\alpha}\pr) v_\alpha\,\,,
\end{equation}
where the external mixing wave function factors are
\begin{eqnarray}
  \label{eq:neutmix}
  {\cal Z}_R^{0\beta \alpha}&=&\frac{
    \mbet\,\SigmaSL{\beta\alpha}(\mas)
    +\ma\,\SigmaSR{\beta\alpha}(\mas)
    +\mbet\,\ma\,\SigmaL{\beta\alpha}(\mas)
    +\mas\,\SigmaR{\beta\alpha}(\mas)                       
  }{\mas-\mbets}\,\,, \nonumber\\
  {\cal Z}_L^{0\beta \alpha}&=&\frac{\mbet\,\SigmaSR{\beta\alpha}(\mas) 
    +\ma\,\SigmaSL{\beta\alpha}(\mas)
    +\mbet\,\ma\,\SigmaR{\beta\alpha}(\mas)
    +\mas\,\SigmaL{\beta\alpha}(\mas)
  }{\mas-\mbets}\,\,.
\end{eqnarray}
The same expression is valid for the creation of 
{a $\cplus_i$ (anti-$\cmin_i$, using ${\cal Z}^{-ji}_{L,R}$)}
changing the indices $\alpha\to i$, $\beta\to j$.\footnote{
The corresponding ones for 
{$\cmin_i$ (anti-$\cplus_i$) are}
${\cal  Z}^{+ji}_{L}\equiv{\cal Z}^{-ji}_{R}$, 
${\cal Z}^{+ji}_{R}\equiv{\cal Z}^{-ji}_{L}$.
}

Admittedly, the choice of the two masses as input
parameters is not ideal. For each pair of on-shell masses $\mco$, $\mct$
there exists up to a fourth-fold discrete ambiguity in the determination
of  the underlying parameters $M$ and $\mu$. There are, however, ways to
determine uniquely these parameters. For example, assuming
one knows both masses, either from a threshold scan in a LC or from the
LHC, one can measure at the LC the production cross-section
$\sigma(e^-e^+\to\cmin_1\cplus_1)$ with different polarizations of the
initial state electron and positron~\cite{Kneur:1999gy}. One finds in this way
experimental values for the mixing matrix elements $U_{ij}^{exp}$ and
$V_{ij}^{exp}$ which disentangle the ambiguity. One uses then the
derived $M$ and $\mu$ parameters to compute the renormalized $U$ and $V$
matrices, which are used for the computation of the radiative corrections. 

A comment is in order regarding other renormalization
prescriptions. Note that, at variance with
Ref.~\cite{Eberl:2001eu,Fritzsche:2002bi}, our 
renormalization prescription does not introduce counterterms for the
mixing matrices $U$, $V$ and $N$. One can, in fact, introduce these
counterterms and fix them in different ways. For example one could
take the point of view that the mixing matrices are functions of the
parameters in the mass matrix:
$$
U=F_1(M,\mu,\mw,\tb) \ \ , \ \ V=F_2(M,\mu,\mw,\tb) \ \ , \ \
N=F_3(M',M,\mw,\mz,\tb)\,\,, 
$$
and then one computes the counterterms as functions of the counterterms
of the mass matrix:
$$
\begin{array}{lcl}
\delta U&=&f_1(\delta M,\delta \mu,\delta \mw,\delta \tb) \ \ , \ \ \\
\delta V&=&f_2(\delta M,\delta \mu,\delta \mw,\delta \tb) \ \ , \ \ \\
\delta N&=&f_3(\delta M',\delta M,\delta \mw,\delta \mz,\delta \tb)\,\,. 
\end{array}
$$
The problem with this approach is that, while the analytic form of the
chargino functions $F_1$ and 
$F_2$ are known, the neutralino function $F_3$ is usually computed
numerically, and then the computation of $f_3$ is not possible. 
We have checked that this renormalization
framework gives exactly the same results {as} 
the non-introduction of 
mixing matrix counterterms for the one-loop partial
decay widths of sfermions into charginos. The authors of
Ref.~\cite{Eberl:2001eu} take a different approach, introducing
independent renormalization conditions for the counterterms of the
mixing matrices. {In Ref.~\cite{Fritzsche:2002bi} the
  counterterms to the $U$, $V$ and $N$ {matrices} are related to
  those of 
  the mixing self-energies\footnote{{See
      Ref.~\cite{Fritzsche:2002bi} for 
  a comparison of the different renormalization schemes.}}.}
 When comparing the results presented here with the ones
of Ref.~\cite{Eberl:2001eu,Fritzsche:2002bi}, one should therefore take into
account that the meaning of the renormalized parameters $M$, $\mu$, $M'$
(and the mixing matrices) is not the same. When comparing physical
quantities (such as pole masses), the results should be equivalent at
one-loop order.

\subsection{Vertex renormalization and decay amplitudes}
\label{sec:renorlagrangian}

Using the notation introduced in the above sections, the tree-level
interaction Lagrangian between fermion-sfermion-(chargino or neutralino)
reads~\cite{Coarasa:1996qa}\footnote{Note, however, a change of
  conventions, in the
  neutralino mass-matrices~(\ref{eq:mmassaneut}), and in the neutralino
  couplings. The change in the couplings allows for a joint presentation
  {of} the chargino and neutralino expressions.}
    \begin{eqnarray}
      \label{LcqsqLR}
      \label{eq:Lqsqcn}
      {\cal L}_{\chi \tilde{f} f'}&=&\sum_{a=1,2}\sum_{r} {\cal L}_{\chi_r
      \tilde{f}_a f'} + \mbox{ h.c.}\,\,,\nonumber\\
     {\cal L}_{\chi_r \tilde{f}_a f'}&=& -g\,\tilde{f}_a^* \bar{\chi}_r
      \left(A_{+ar}^{(f)}\pl +  A_{-ar}^{(f)}\pl\right) f'\,\,.
    \end{eqnarray}
Here we have adopted a compact notation, where
$f'$ is either $f$ or its 
$SU(2)_L$ partner for $\chi_r$ being a neutralino or a chargino,
respectively. Roman characters 
   $a,b\ldots$ are reserved for sfermion indices and 
   $i,j,\ldots$ for chargino indices; 
   Greek indices $\alpha,\beta,\ldots$ denote neutralinos;
   Roman indices $r,s\ldots$ indicate either a chargino or a
   neutralino. For example, the top-squark interactions with charginos
   are obtained by replacing $f\to t$, $f'\to b$, $\chi_r\to \cmin_r$,
   $r=1,2$. The coupling matrices that encode the dynamics are given by
    \begin{eqnarray}
        \label{V1Apm}
        \Apit &=& \Rot\Vo^*-\lt\Rtt\Vt^*\, ,\nonumber\\
        \Amit &=& -\lb\Rot\Ut\, ,\nonumber\\
        \Apat &=&\frac{1}{\sqrt{2}} \left(
            \Rot\left(\Nt^*+\YL\tw\No^*\right)
            +\sqrt{2}\lt\Rtt\Nf^*
          \right)\, ,\nonumber\\
        \Amat &=& \frac{1}{\sqrt{2}} \left(
            \sqrt{2}\lt\Rot\Nf
            -\YRt\tw\Rtt\No
            \right)\, ,\nonumber\\
        \Apib &=& \Rob\Uo^*-\lb\Rtb\Ut^*\, ,\nonumber\\
        \Amib &=& -\lt\Rob\Vt\, ,\nonumber\\
        \Apab &=& -\frac{1}{\sqrt{2}} \left(
          \Rob\left(\Nt^*-\YL\tw\No^*\right)
          -\sqrt{2}\lb\Rtb\Nth^*
        \right)\, ,\nonumber\\
        \Amab &=& -\frac{1}{\sqrt{2}} \left(
          -\sqrt{2}\lb\Rob\Nth
          +\YRb\tw\Rtb\No
          \right) \, ,
     \end{eqnarray}
with $\YL$ and $Y_R^{t,b}$ the weak hypercharges of the left-handed
$SU(2)_L$ doublet and right-handed singlet fermion, and
$\lt=\mt/(\sqrt{2}\mw\sin\beta)$ and $\lb=\mb/(\sqrt{2}\mw\cos\beta)$
are the Yukawa couplings normalized to the $SU(2)_L$ gauge coupling
constant $g$.

As far as vertex renormalization is concerned, the vertex counterterms are
already determined by the renormalization procedure described above.
Introducing the one-loop counterterms analogously
to~\cite{Guasch:1998as} we obtain the following counterterm
Lagrangian~\cite{Guasch:1998as,Coarasa:1996qa}
\begin{eqnarray}
  \delta {\cal L}_{\chi_r \sfra f'}&\equiv&  
  g
    \sfra^{*}\bar{\chi_r}\left(\delta\LApaf\PL+\delta\LAmaf\PR\right)\,f'
    +\mbox{\rm h.c.}\nonumber\\
&=&   \frac{1}{2}\left\{
  \left[\dalpha+\frac{\cwp^2}{\swp^2} \left(\dmws-\dmzs\right)\right]
    + \dz_{\sfr}^{a}\right\}  {\cal L}_{\chi_r  \sfra f'} 
  +{\cal L}_{\chi_r \sfrb f' }\dz_{\sfr}^{ba}\nonumber\\
  &&+\Bigg\{{g}\sfra^{*}\bar{\chi_r}\left[\Apaf
  \frac{1}{2} \left(\Dzrchir+\DZlf\right) \PL
  +\Amaf \frac{1}{2} \left(\Dzlchir+\DZrf\right)\PR\right]\,f'\nonumber\\
  &&+{g}\sfra^{*} \bar{\chi}_r \left(\delta\Apaf
    \PL+\delta \Amaf \PR\right)\,f'
  +\mbox{ h.c.}\Bigg\}\,\,,\,\,(b\neq a)\,\,,\nonumber \\
%% Counterterms \delta A
        \delta \Apit &=&
        \delta\Rot\Vo^*-\left(\lt\delta\Rtt+\delta
        \lt\Rtt \right)\Vt^*\, ,\nonumber\\ 
        \delta\Amit &=& -( \lb\delta\Rot +\Rot \delta \lb )\Ut\, ,\nonumber\\
        \delta\Apat &=&\frac{1}{\sqrt{2}} \left(
            \delta\Rot\left(\Nt^*+\YL\tw\No^*\right)
            +\Rot \YL\No^*\delta\tw
            +\sqrt{2}\left(\lt\delta\Rtt+\Rtt\delta\lt\right)\Nf^*
          \right)\, ,\nonumber\\
        \delta\Amat &=& \frac{1}{\sqrt{2}} \left(
            \sqrt{2}(\lt\delta\Rot+\Rot\delta\lt)\Nf
            -\YRt(\tw\delta\Rtt+\Rtt\delta\tw)\No
            \right)\, ,\nonumber\\
        \delta\Apib &=&
        \delta\Rob\Uo^*-(\lb\delta\Rtb+\Rtb\delta\lb)\Ut^*\,
        ,\nonumber\\ 
        \delta\Amib &=& -(\lt\delta\Rob+\Rob\delta\lt)\Vt\, ,\nonumber\\
        \delta\Apab &=& -\frac{1}{\sqrt{2}} \left(
          \delta\Rob\left(\Nt^*-\YL\tw\No^*\right)
          -\Rob \YL\No^*\delta\tw
          -\sqrt{2}(\lb\delta\Rtb+\Rtb\delta\lb)\Nth^*
        \right)\, ,\nonumber\\
        \delta\Amab &=& -\frac{1}{\sqrt{2}} \left(
          -\sqrt{2}(\lb\delta\Rob+\Rob\delta\lb)\Nth
          +\YRb(\tw\delta\Rtb+\Rtb\delta\tw)\No
          \right)\, .\nonumber\\
  \frac{\delta \lb}{\lb}&=&\dmb-\frac{1}{2}
  \dmws-\Dcosb\,\,,\,\,
  \frac{\delta \lt}{\lt}=\dmt-\frac{1}{2}
  \dmws-\Dsinb\,\,,\,\,\nonumber\\
  \delta{t_W}&=&\frac{1}{2 s_W c_W}\left(\frac{\delta
    M_Z^2}{M_Z^2}-\frac{\delta M_W^2}{M_W^2} \right) ,
  \label{eq:counterl}
\end{eqnarray}
where $\delta\alpha$, $\delta M_{W,Z}^2$ are the charge and mass
counterterms for the MSSM, as given in~\cite{Garcia:1994sb}, and the
counterterms for the mixing matrices 
have been defined in eq.~(\ref{eq:bareweakmass}). 

The renormalized amplitude for the decay
$\sfra \rightarrow f' \chi_r$ 
can then be written at the one-loop level as follows,
\begin{eqnarray}
  -i T_{ar}&\equiv&-i T(\sfra \rightarrow f' \chi_r)=-i T_{ar}^{\rm
  tree} -i T_{ar}^{\rm loop}\,\,,\nonumber\\
-i T_{ar}^{\rm tree}&=&i
   {g}\bar{u}_{f'} \left[\Aprfc\PR
     +\Amrfc\PL\right] v_{\chi_r}\,\,,\nonumber\\
  -i T_{a r}^{\rm loop}&=&i
   {g}\bar{u}_{f'} \left[\Cpaf\PR
     +\Cmaf\PL\right] v_{\chi_r}\,\,;\,\,%\nonumber\\
   \Cpmaf=\delta\LApmafc+\LApmaf{}^{\Sigma}
   +\LApmaf{}^{\rm 1PI}\,\,.
  \label{eq:renormamp} 
\end{eqnarray}
It contains,
besides the counterterms $\delta\Lambda$ from (\ref{eq:counterl}),
the one-loop contributions $\Lambda^{\rm 1PI}$ 
to the one-particle-irreducible three-point vertex functions 
{shown in Fig.~\ref{diags:3point}a},
and the quantities
$\Lambda^{\Sigma}$ 
corresponding to the
higher-order terms from the two-point functions
in the one-loop expansion of the general
expression~(\ref{eq:externalZ}), explicitly given by
\begin{equation}
\LApaf{}^{\Sigma}=\sum_{s\neq r} \Apsfc {\cal Z}_R^{sr} \ \ ,\ \ 
\LAmaf{}^{\Sigma}=\sum_{s \neq r} \Amsfc {\cal Z}_L^{sr}\,\,,
  \label{eq:mixingeffect}
\end{equation}
where ${\cal Z}_{\{L,R\}}^{sr}$ 
has been defined in~(\ref{eq:neutmix}).

\figtwo

We are now ready to compute the partial decay widths. 
The tree-level expressions read
\begin{eqnarray}
  \Gamma^0_{ar}&=&\Gamma^0(\sfra \to f' \chi_r)=
  \frac{g^2}{16\,\pi\,\msfa^3}\,\lambda
  (\msfas,M_r^2,m_{f'}^2)\times\,  \nonumber\\ 
  &&\times\left[
    (\msfas-M_r^2-m_{f'}^2)
    \left(|A^{(f)}_{+ar}|^2+|A^{(f)}_{-ar}|^2 \right)
    -4\,m_{f'}\,M_r {\rm Re}\left(A^{(f)}_{+ar}\,A^{(f)*}_{-ar}\right)
  \right]\,\,,
  \label{eq:treleevelgamma}
\end{eqnarray}
with $\lambda(x^2,y^2,z^2)=\sqrt{ [x^2-(y-z)^2][x^2-(y+z)^2]}$.

Due to the presence of photon loops, the one-loop partial decay width
computed using the amplitude~(\ref{eq:renormamp})
is infrared divergent;
hence, bremsstrahlung of real photons has to be added  
to cancel {this divergence}. 
We therefore
include in our results the radiative partial
decay width $\Gamma(\sfra \to f' \chi_r \gamma)$, 
including both the soft and the hard photon part.\footnote{Except for
  the partial decay width $\Gamma(\sneut\to\nu\neut_\alpha \gamma)$,
  which is obviously zero at tree-level.}
{The corresponding Feynman diagrams are shown in
  Fig.~\ref{diags:3point}b}. 
This yields finally
the complete one-loop electroweak correction,
\begin{eqnarray}
  \delta^{ar}&=&
  \displaystyle\frac{\Gamma(\sfra\to f' \chi_r)}{\Gamma^0(\sfra\to f'\chi_r)}-1
 = \delta^{ar}_{ {\rm virt}}+\frac{\Gamma(\sfra \to f' \chi_r \gamma)
}{\Gamma^0(\sfra\to f'\chi_r)}\,\,,\nonumber\\ 
\displaystyle \delta^{ar}_{{\rm virt}}&=& \displaystyle
  2{\rm Re}\left[  (\msfas-M_r^{2}-\mfps) (\Apaf \Cpaf + \Amaf
      \Cmaf)\right. \nonumber\\
&&\left.\ \ \ 
 - 2 M_r \mfp (\Apaf \Cmaf+\Amaf
   \Cpaf) \right] 
 \times \nonumber \\
&& \left[(\msfas-M_r^{2}-\mfps) (|\Apaf|^2+|\Amaf|^2) -4 M_r \mfp {\rm
   Re}(\Apaf \Amafc)\right]^{-1}\,\,.
  \label{eq:deltadef}
\end{eqnarray}

The loop computation itself is rather tedious, since there is a huge
number of diagrams to compute. This is better done by
means of automatized tools.
The computation of the loop diagrams has been 
performed
by using the Computer
Algebra Systems \textit{FeynArts 3} and
\textit{FormCalc 2.2}~\cite{Kublbeck:1990xc,Hahn:1998yk}. 
We have
produced a set of Computer Algebra programs that compute the one-loop
diagrams (and the bremsstrahlung corrections), which are then plugged into
a \textit{Fortran} code for the numerical evaluation with the help of the
one-loop routines
\textit{LoopTools 1.2}~\cite{Hahn:1998yk}. 
A number of checks {have} been made  on the results. 
The UV and infra-red
finiteness of the result, relying on the relations between the
different sectors of the model, is a non-trivial check. We also have
recovered results already available in the
literature; for instance, we used our set of programs to reproduce the strong
corrections of~\cite{Djouadi:1997wt}, and, using the \textit{higgsino}
approximation, we could also reproduce the results
of~\cite{Guasch:1998as}. Moreover we also checked that, when using the
$\overline{\mbox{MS}}$-scheme, the one-loop corrections to neutralino
and chargino masses reproduce those of~\cite{Pierce:1994gj}. 

Although we consider the chargino and neutralino masses as input
parameters, in our numerical study we treat them in a slightly different
way. We choose a set of renormalized input parameters $(M,M',\mu)$, and
apply~(\ref{eq:mmassaneut}), (\ref{eq:defUVN}) to obtain the one-loop
renormalized masses. 
Of course, if SUSY would be discovered the procedure will
be the other way around, that is, the MSSM parameters will be computed
from the various observables measured, for example, from the chargino
production cross-section and asymmetries at 
the LC~\cite{Kneur:1999gy}. 
For a consistent treatment, the one-loop expressions for
these observables will have to be used~\cite{Blank:2000uc}.

\subsection{Universal corrections: Non-decoupling effects and
  effective coupling matrices}
\label{sec:universal}

We note that  there exists a certain combination of contributions in the
one-loop amplitude~(\ref{eq:renormamp}) that does not depend on
the sfermion flavour. These 
contributions can      be expressed formally as corrections to the coupling
matrices
\begin{equation}
  \label{eq:effective1}
    \tilde{U}=U+\Delta U \ \ , \ \ 
    \tilde{V}=V+\Delta V  \ \ ,  \ \  \tilde{N}=N+\Delta N  \ \ , 
\end{equation}
with
\begin{eqnarray}
  \Delta U_{i1}&\equiv& U_{i1}
  \,\left(\dgog+\frac{\Dzrcharm{}^i}{2}\right) +U_{j1} {\cal Z}_R^{-ji}\,\,,
\nonumber\\
  \Delta U_{i2}&\equiv& U_{i2}\left(\dgog+\frac{\Dzrcharm{}^i}{2}-\frac{1}{2}\dmws -\Dcosb\right)+U_{j2} {\cal Z}_R^{-ji}\,\,,\nonumber\\
  \Delta V_{i1}&\equiv& V_{i1} \,\left(\dgog+\frac{\Dzlcharm{}^i}{2}\right) +V_{j1} {\cal Z}_L^{-ji}\,\,,\nonumber\\
  \Delta V_{i2}&\equiv& V_{i2}\left(\dgog+\frac{\Dzlcharm{}^i}{2}-\frac{1}{2}\dmws -\Dsinb\right)+V_{j2} {\cal Z}_L^{-ji}\,\,,\nonumber\\
  \Delta N_{\alpha1}&\equiv&N_{\alpha1}
    \left(\dgog
    +\frac{\Dzrneut{}^\alpha}{2}+\dtow\right)+\sum_{\beta\neq\alpha} N_{\beta 1} {\cal Z}_R^{0\beta\alpha}\,\,,\nonumber\\
  \Delta N_{\alpha2}&\equiv&  N_{\alpha2}
\left(\dgog
    +\frac{\Dzrneut{}^\alpha}{2}\right)+\sum_{\beta\neq\alpha} N_{\beta 2} {\cal Z}_R^{0\beta\alpha}\,\,,\nonumber\\
  \Delta N_{\alpha3}&\equiv& N_{\alpha3}
  \left(\dgog+\frac{\Dzrneut{}^\alpha}{2}-\frac{1}{2}\dmws
  -\Dcosb\right)+
\sum_{\beta\neq\alpha} N_{\beta3} {\cal Z}_R^{0\beta\alpha}\,\,,\nonumber\\
  \Delta N_{\alpha4}&\equiv&
  N_{\alpha4}\left(\dgog+\frac{\Dzrneut{}^\alpha}{2}-\frac{1}{2}\dmws
  -\Dsinb\right)+
\sum_{\beta\neq\alpha} N_{\beta4}
  {\cal Z}_R^{0\beta\alpha}\,\,,  \nonumber\\
\frac{\delta g}{g}&\equiv  &\frac{1}{2}\,\left(\dalpha+\frac{\cwp^2}{\swp^2}\,\left(\dmws-\dmzs\right)\right)\,\,.
\label{eq:universal} 
\end{eqnarray}
Unfortunately the full
contributions to the expressions~(\ref{eq:universal}) are
divergent. The only consistent subset of corrections which makes all the
expressions in~(\ref{eq:universal}) finite is the subset of fermion and
sfermion 
loops {contributing to the self-energies of the gauge bosons,
 Higgs bosons, charginos and neutralinos}. With this restriction, we can
define  \textit{effective coupling   matrices} 
\begin{equation}
  \label{eq:effectivegen}
  U^{eff}=U+\Delta U^{(f)} \ \ , \ \ 
  V^{eff}=V+\Delta V^{(f)}  \ \ ,  \ \  N^{eff}=N+\Delta N^{(f)}  \ \ , 
\end{equation}
where $\Delta U^{(f)}$, $\Delta V^{(f)}$, $\Delta N^{(f)}$ are given by
the expressions 
(\ref{eq:universal}) taking into account only loops of fermions and
sfermions. We will refer to these corrections as \textit{universal
  corrections}. They are the equivalent of the
  \textit{super-oblique corrections} of Ref.\cite{Katz:1998br}.

These effective coupling matrices present a very interesting
feature. Since the divergences in the SM sector of the model (gauge and
Higgs sectors) cancel 
the divergences of the chargino-neutralino sector, the associated logarithmic
terms ($\log(m/\mu^D)$, $\mu^D$ being 
the arbitrary mass parameter of dimensional reduced integrals) must be
combined. As a result, a non-decoupling term $\sim\log(m^{SUSY}/m^{SM})$
appears in the final expression. Here $m^{SUSY}$ represents a generic
SUSY mass, and $m^{SM}$ a generic SM mass. 

We have checked explicitly this effect. We have computed analytically
the electron-selectron contributions to the $\Delta U$ and $\Delta V$
matrices~(\ref{eq:universal}), assuming zero mixing angle in the
selectron sector ($\theta_e=0$), we have identified the leading terms in the
approximation $m_{\tilde e_i}, m_{\tilde \nu}\gg (\mw,\mi) \gg m_e$, and
 analytically canceled the divergences and the $\log(\mu^D)$ terms;
finally, we have kept only the terms logarithmic in the {slepton}
masses. The result reads as follows:
\begin{eqnarray}
\Delta\UChaio^{(f)}&=& \commonnumfactor\log\left(\frac{M^2_{\tilde e_L}}{\mxs}\right)\,\bigg[
 \frac{\UChaio^3}{6} - 
  \UChait \frac{\sqrt{2}\,\mw\,(M\,\cbta +
  \mu\,\sbta)}{3\,(M^2-\mu^2)\,\mchasomenysmchast^2} %\times \nonumber\\&&
\left(M^4 - M^2\,\mu^2 + \right.\nonumber\\
&&\left. + 3\,M^2\,\mws + 
     \mu^2\,\mws + \mwf + \mwf\,\cfbt %\right.\nonumber\\&&
    + (\mu^2-M^2)\,\mis + 
     4\,M\,\mu\,\mws\,\stbt\right)\,\bigg]\,\,,
\nonumber\\
\Delta\UChait^{(f)}&=&\commonnumfactor\log\left(\frac{M^2_{\tilde e_L}}{\mxs}\right)\,
\UChaio\,\frac{\mw\,(M\,\cbta + \mu\,\sbta)}
 {3\,\sqrt{2}\,(M^2-\mu^2)\,\mchasomenysmchast^2} \times \nonumber\\
  &\times&  \left((M^2-\mu^2)^2 
+ 4\,M^2\,\mws + 4\,\mu^2\,\mws + 2\,\mwf + 
   2\,\mwf\,\cfbt + 8\,M\,\mu\,\mws\,\stbt\right)\,\,,
\nonumber\\
\Delta\VChaio^{(f)}&=&\commonnumfactor\log\left(\frac{M^2_{\tilde e_L}}{\mxs}\right)\,\bigg[
\frac{\VChaio^3}{6}
- 
 \VChait \frac{\sqrt{2}\,\mw\,(\mu\,\cbta +
  M\,\sbta)}{3\,(M^2-\mu^2)\,\mchasomenysmchast^2} %\times \nonumber\\&&
\left(M^4 - M^2\,\mu^2 +\right.\nonumber\\
&&\left. + 3\,M^2\,\mws + 
     \mu^2\,\mws + \mwf + \mwf\,\cfbt 
 + (\mu^2-M^2)\,\mis + 
     4\,M\,\mu\,\mws\,\stbt\right)\bigg]\,\,,
\nonumber\\
\Delta\VChait^{(f)}&=&\commonnumfactor\log\left(\frac{M^2_{\tilde e_L}}{\mxs}\right)\,\VChaio\,
\frac{\mw\,(\mu\,\cbta + M\,\sbta)}
 {3\,\sqrt{2}\,(M^2-\mu^2)\,\mchasomenysmchast^2} \times \nonumber\\
  &\times&  \left((M^2-\mu^2)^2  
   + 4\,M^2\,\mws + 4\,\mu^2\,\mws + 2\,\mwf + 
   2\,\mwf\,\cfbt + 8\,M\,\mu\,\mws\,\stbt\right),\nonumber\\
\label{eq:logterms}
\end{eqnarray}
$M^2_{\tilde e_L}$ being the soft-SUSY-breaking mass of the
$(\tilde{e}_L,\tilde{\nu})$ doublet~(\ref{eq:sbottommatrix}),
whereas $\mx$ is a SM mass.
{In the on-shell scheme for the SM electroweak theory we define
parameters at very different scales, basically $\mx=\mw$ and
$\mx=m_e$. These wide-ranging scales enter the structure of the
counterterms -- see the last formula in eq.(\ref{eq:universal})-- and so
must appear in eq.(\ref{eq:logterms}) too. As a result the leading log
in the various terms of this equation will vary accordingly. For
simplicity in the notation we have factorized $\log M^2_{\tilde
e_L}/\mxs$ as an overall factor. In some cases this factor can be very
big, $\log M^2_{\tilde e_L}/m_e^2$; it comes from the electron-selectron
contribution to the chargino-neutralino self-energies. Its
non-decoupling behaviour is logarithmic in the heavy (SUSY) mass and it
can be explained from renormalization group arguments relating the
supersymmetric gauge couplings at the SM and SUSY scales (see
below). This term is similar to the logarithmic part of the universal
effects from the SM gauge bosons, which is related to the
renormalization of the ordinary gauge couplings and of course can also
be explained by renormalization group arguments -- in this case
involving the internal SM scales $\mw$ and $m_e$.}
Equivalent
expressions can be found for the quark-squark sector. We have checked
numerically 
that the expressions~(\ref{eq:logterms}) approximate well the
logarithmic term in the full expression~(\ref{eq:effectivegen}). We have
computed numerically~(\ref{eq:effectivegen}) for different input parameters
(including $\theta_e\neq0$) 
for selectron masses in the range $m_{\tilde e_1}=1-100\TeV$, we  have
fitted this numerical 
results to a simple function 
$f(m_{\tilde{e_1}})=A+B \log(m_{\tilde{e_1}})$. The results show a
correlation factor close to $\pm1$ in better than $10^{-4}$. The simple
expressions~(\ref{eq:logterms}) approximate the coefficient of the
logarithmic term in better than $1\%$. 
\figthree

However, the expressions~(\ref{eq:logterms}) do not reproduce the full
result~(\ref{eq:effectivegen}) due to the presence of important
non-logarithmic terms. Upon adding up the three slepton generations,
the contributions to~(\ref{eq:effectivegen}) can be typically of the order
of $\sim3\%$ for $m_{\tilde e_1}=1\TeV$, whereas the
approximation~(\ref{eq:logterms}) gives $\sim 2\%$ under the same
conditions. 

We want to stress that this is a physical, measurable, effect. By
measuring the two chargino masses ($M_1$, $M_2$), one can extract the
SUSY parameters $M$ and $|\mu|$, and by assuming CP-conservation one can
obtain the renormalized mixing matrices $U$ and $V$ for each sign of
$\mu$. {On the other hand,} 
one can extract the value of the mixing matrices in a
polarized $e^+e^-$ linear collider~\cite{Kneur:1999gy}. However, the
extracted values of the mixing matrices are, in a first approximation,
the ones of eq.~(\ref{eq:effectivegen}), which can deviate from the
on-shell ones --eq.~(\ref{eq:defUVN})-- at the several percent level. In
this way, even if some 
(or all) of the sfermions have masses beyond the reach of an $e^+e^-$ linear
collider, one can get information of their mass scale by means of the
effective coupling matrices. In fact, the larger the mass,
  the larger is the correction.  Of course, the experimental
value of the 
cross-section $\sigma(e^+e^-\to \cplus\cmin)$ will have to be compared
with the full one-loop computed cross-section~\cite{Blank:2000uc}, since
the rest of the one-loop corrections can be as large the
{contribution} 
of the effective coupling matrices~(\ref{eq:effectivegen}).

The {ultimate} reason for these non-decoupling effects lies in the
  breaking of SUSY, which affects  the SUSY relation
  between the gauge-boson and gaugino couplings (or Higgs-boson and
  higgsino couplings). 
  As a simple example, SUSY {implies} 
that the 
  $e^+e^-\gamma$ coupling must be equal to the 
  $\tilde{e}^+e^-\tilde{\gamma}$ coupling. 
  For broken SUSY, this
  equality is lost, and the deviation of the
  $\tilde{e}^+e^-\tilde{\gamma}$-coupling from the
  $e^+e^-\gamma$-coupling  grows with the scale of the
  SUSY breaking~\cite{Katz:1998br,Hikasa:1996bw}. 
{One can understand the appearance of the non-decoupling effects
  by renormalization group arguments. In an energy scale much larger
  than any SUSY mass scale ($Q\gg m^{SUSY}$) the theory is
  supersymmetric, and the effective $e^+e^-\gamma$ gauge coupling
  ($\alpha(Q)$) is equal to the effective $\tilde{e}^+e^-\tilde{\gamma}$
  Yukawa coupling ($\tilde{\alpha}(Q)$), and their renormalization group
  equations (RGE) are the same. If some hierarchy exists in the SUSY
  sector (say, for definiteness $m_{\tilde q}>m_{\tilde e}$), at the
  scale $Q=m_{\tilde q}$ the squarks decouple from the running of
  $\alpha$ and $\tilde{\alpha}$. Quarks, on the other hand, decouple from
  $\tilde{\alpha}$ but not from $\alpha$. Therefore, at scales 
  $Q<m_{\tilde  q}$, {the coupling} $\tilde{\alpha}$ is frozen at the squark mass scale
  $\tilde{\alpha}(Q<m_{\tilde q})=\tilde{\alpha}(m_{\tilde q})=\alpha(m_{\tilde q})$
  as far as quark/squark contributions are
  concerned\footnote{{Of course, 
  there is an evolution of $\tilde{\alpha}(Q<m_{\tilde q})$ due to the
  lepton/slepton contributions. Here we leave the slepton contributions
  out of the discussion for the sake of clarity.}}. Therefore, the
  comparison between the two couplings gives, at one-loop order:}
$$
\frac{\tilde{\alpha}(Q)}{\alpha(Q)}-1=\frac{\alpha(m_{\tilde
    q})}{\alpha(Q)}-1 = \beta
\log \frac{m_{\tilde q}}{Q} \ \ ,\ \  Q<m_{\tilde q}\,\,,
$$
{where the QED $\beta$-function does not include the contribution
   from squarks.}
{Since we are using an on-shell renormalization scheme, we are
  comparing in fact $\tilde{\alpha}(Q)/\alpha(0)$, and since the quarks
  decouple from $\alpha$ at $Q=m_{q}$, {we  end up} with a
  correction proportional to $\log m_{\tilde q}/m_{q}$. We have
  explicitly checked this fact using SUSY Quantum Electrodynamics as a
  toy model. The 
  complete electroweak model looks much cumbersome, since various
  quantities are fixed at different scales (e.g. the  masses of
  the gauge bosons $\mw$, $\mz$ are fixed at their respective pole
  values, and the electromagnetic constant is fixed in the Compton limit
  $\alpha(0)$), therefore different pieces of the corrections carry
  different {scales} $Q$ {in} the arguments of the logarithms.}

\figfour

In Figs.~\ref{fig:Ueff} and~\ref{fig:Veff} we show contour plots in the
$M-\mu$ plane of the relative correction to the elements of the mixing matrices
$U$ and $V$~(\ref{eq:effectivegen}). For this figure we have chosen a sfermion
spectrum around $1\TeV$, namely:
\begin{equation}
  \label{eq:numbersUV}
\begin{array}{l}
\tb=4\,\,,\,\,m_{\tilde{l}_2}=m_{\tilde{d}_2}=m_{\tilde{u}_2}=1 \TeV \,\,,\,\,
 m_{\tilde{l}_1}=m_{\tilde{d}_1}=m_{\tilde{l}_2}+5\GeV\,\,,\,\,\\
\theta_l=\theta_q=\theta_b=0 \,\,,\,\, \theta_t=-\pi/5\,\,,
\end{array}
\end{equation}
where we have assumed a common mass for all the charged sleptons and
down-type quarks, and considered mixing only in the  stop
sector.

The thick black lines in Figs.~\ref{fig:Ueff}-\ref{fig:Veff} correspond
to {spurious} divergences in the relative corrections. The ones
corresponding to 
$M=\pm \mu$ are divergences in the corrections themselves, since our
renormalization prescription fails in this case --see
eq.~(\ref{eq:soldM2dmu}). This divergence appears also explicitly in the
approximate expression~(\ref{eq:logterms}) as the inverse of
$M^2-\mu^2$. The other divergences correspond to lines where the
renormalized mixing matrix elements are zero. These correspond to
$M=-\mu\tb$ ($M=-\mu /\tb$) for $U_{11}$, $U_{22}$ ($V_{12}$,
$V_{21}$). Corrections as large as $\pm10\%$ can only be found in the
vicinity of these divergence lines. However, there exist large regions
of the $\mu-M$ plane  where the corrections are larger than $2\%$,
$3\%$, or even $4\%$.

\section{Numerical evaluation}
\label{sec:numeric}

{In this section we tackle the (cumbersome) numerical analysis of the
corrections to the various $\sfr\to f'\chi$ partial decay channels
according to the following plan. First of all we assess the relevance of
the universal corrections defined in sect.~\ref{sec:universal}. Next we
focus on the non-universal corrections, with especial emphasis on the
non-decoupling effects from gauginos and Higgs particles, followed by an
exhaustive analysis of the corrections to the various partial decay
widths as a function of the most relevant parameters. After that, we
briefly concentrate on the strong corrections. Finally, we combine the
universal, non-universal, and strong corrections to the partial decay
widths to evaluate the impact on the branching ratios  for all the decay
channels, which are in practice the true observables.}

As for the presentation of the results themselves, we will use the
following criteria: The first and second generation of sfermions have
very similar properties, so only the results for the first generation
are presented. We will present separate results for the third squark
generation ($\stopp,\sbottom$), since the large Yukawa couplings and
masses make this generation  behave in a special way. The third
generation sleptons ($\stau,\tilde{\nu}_\tau$) have also large Yukawa
couplings, but their effect is small, unless \tb\  is very large. We
will refrain 
to show {explicit} results {for}
$\stau,\tilde{\nu}_\tau$, {if} these 
results are similar to the first generation ones.

We have used the following {default set of central parameters} for our
numerical evaluation:
\begin{equation}
\begin{array}{l}
 \mt=175\GeV\,\,,
 \mb=5\GeV\,\,,
 \tb      =  4\,\,,
 \mHp  =  120\GeV\,\,,\\
 \msbt=\msdt=\msut=\mselt=300\GeV\,\,,\\
 \msbo=\msdo=\mselo=\msbt+5\GeV\,\,,
\msut=290\GeV\,\,,
 \mstt=300\GeV\,\,,\\
 \osb=\osd=\osu=\osel=0\,\,,
 \ost=-\pi/5\,\,,\\
 \mu      =  150\GeV\,\,,
 M       =  250\GeV\,\,.
\end{array}
\label{eq:inputpars}
\end{equation}
{We point out that the amount
 of splitting chosen in the sbottom sector is not critical for the
 numerical results, and has nothing to do with preserving the vacuum
 condition (\ref{eq:necessary}) because we are assuming zero sbottom
 mixing angle in the first eq.(\ref{eq:Abt}). Furthermore, the negative
 sign for the stop mixing angle is related to the chosen sign for $\mu$
 and the desired relation $\mu A_t<0$ via the second eq.(\ref{eq:Abt}).}
The other SM parameters have been taken from Ref.\cite{PDB2000}. For
simplicity, we will be using the Grand Unification (GUT) relation
between the electroweak soft-SUSY-breaking gaugino masses
$M'=5\,M\,\tw^2/3$ unless stated otherwise.
The computed values of the heaviest up-type sfermions are given in
table~\ref{tab:sup2masses}.
In
table~\ref{tab:massbr} we show the
tree-level chargino and neutralino masses, as well as the 
{tree-level 
branching} ratios of sfermions decaying into charginos and neutralinos,
and the one-loop corrections to the neutralino masses.

\tableone

\tabletwo

Given the large number of decay channels, a presentation including 
{the
corrections to each individual final state} 
would be tedious. Moreover, a large number of channels have
a very small branching ratio, and 
higher-order terms are not of 
phenomenological interest. For these reasons most of the 
discussion will be devoted to 
the total decay widths of sfermions into charginos and
neutralinos, 
in particular to the relative correction
\begin{equation}
  \label{eq:totalcn}
\delta(\tilde{f_a}\to f'\chi)=\frac{ \sum_r 
    \left(\Gamma(\tilde{f_a}\to  f'\chi_r)-\Gamma^0(\tilde{f_a}\to  f'\chi_r)\right)}
  {\sum_r \Gamma^0(\tilde{f_a}\to  f'\chi_r)}\,\,, 
\end{equation}
with $\chi=\chi^\pm$ or $\chi=\neut$. 
We will not show results for processes {whose} branching ratio
{are} less that 10\% in all of the explored parameter space. 

\subsection{Universal Corrections}
\figfive

\figsix

{As we have said, we start
our numerical analysis by testing the non-decoupling effects directly
associated to the universal corrections discussed in
section~\ref{sec:universal}.}
The main aim of this section is to assess
the numerical impact of these non-decoupling effects. 
To this
end we present in Fig.~\ref{fig:unilep} the universal corrections to the
partial decay widths of the first generation of sleptons
($\tilde{e}-\tilde{\nu}_{e}$) as a function of a common mass for all
 squarks. Since these corrections are universal, and the lepton
masses can be 
safely neglected, they are the same 
for the other generations of sleptons. Looking at the right end of the plots
($m_{\tilde{q}}>10^3\GeV$), the logarithmic scale of the plots makes
evident the 
non-decoupling of squarks by means of a logarithmic term $\log(m_{\tilde{q}})$
equivalent to that of eqs.~(\ref{eq:logterms}). The corrections are also
non-negligible for light squark masses. We see that for squark masses
below $1 \TeV$ the corrections reach a $5\%$ value for most of the decay
channels, or even larger than $10\%$ for the selectron decay into
neutralinos. 

We turn now our attention to the squark decays. Fig.~\ref{fig:unisq} shows
the universal corrections to the squark partial decay widths as a
function of a common mass for all sleptons. We show the
corrections for down-type and up-type quarks of the first and third
generation. Again, the logarithmic behaviour from
eq.~(\ref{eq:logterms}) is evident in this figure. 
{The logarithmic regime is attained already for slepton masses of
  order $1\TeV$.}
The universal corrections are seen to be
positive for all squark decays, ranging between $4\%$ and $10\%$ for
slepton masses below $1\TeV$. 

\subsection{Non-Universal corrections}

{The non-universal corrections comprise the set of full
  corrections excluding the universal corrections of
  section~\ref{sec:universal}. For a given sfermion decay, these
  corrections do not depend on the parameters of the sfermions 
  of different type or generation. That is, e.g. for a $\stopp$ decay, the
  dependence on the lepton/slepton parameters and the first and second
  generation quark/squark parameters appears only in the universal
  corrections analyzed in the previous section, and the non-universal
  corrections depend only on the gauge, Higgs, chargino/neutralino and
  stop/sbottom sectors.  }

\subsubsection{Non-decoupling effects}
\label{sec:nouninondec}
\figseven

The non-decoupling of gauginos was already discussed 
in~\cite{Hikasa:1996bw,Djouadi:1997wt}. There, the QCD corrections to
the squark decays 
{were} computed, and an explicit non-decoupling gaugino term (of
the form 
$\log\mg$) was found. The origin of this non-decoupling effect is
similar to that of the sfermionic ones in~(\ref{eq:logterms}), namely the
gaugino UV-divergences cancel the gauge boson ones, so that the
logarithms associated with the divergence must compensate between SUSY
and non-SUSY particles, {leading to a  $\log(M_{gaugino})$ behaviour
 which can again be explained from simple renormalization group
 arguments.} 
The electroweak corrections present several peculiarities that prevent
from obtaining a simple analytical result of these logarithms. First of
all, the non-decoupling particles are part of the final state of the
process, {thus} 
if their masses are large the decay will be phase-space
closed. Second, the complicated structure of the electroweak sector
 involves the mixing of gauginos and higgsinos. One can not compute
simply the gauge--gaugino or the Higgs-boson--higgsino corrections, unless
very special limits are taken~\cite{Guasch:1998as}. Thus, a numerical
approach is preferred in this case. 

In this subsection (\ref{sec:nouninondec}) we give up the GUT assumption between the gaugino
masses, and will let the soft-SUSY-breaking mass parameters $M$ and $M'$
vary independently. In this way we can afford having a light neutralino
(when $M'$ is light), while the $SU(2)$ gauginos are heavy, and analyze
the non-decoupling effects of the $SU(2)$ gauginos in the neutralino
decays. On the other hand, by maintaining $M$ light, but taking $M'$ to
be heavy, one obtains a heavy $U(1)_Y$ neutralino whose non-decoupling
effects can be studied in the chargino sector.

\figeight

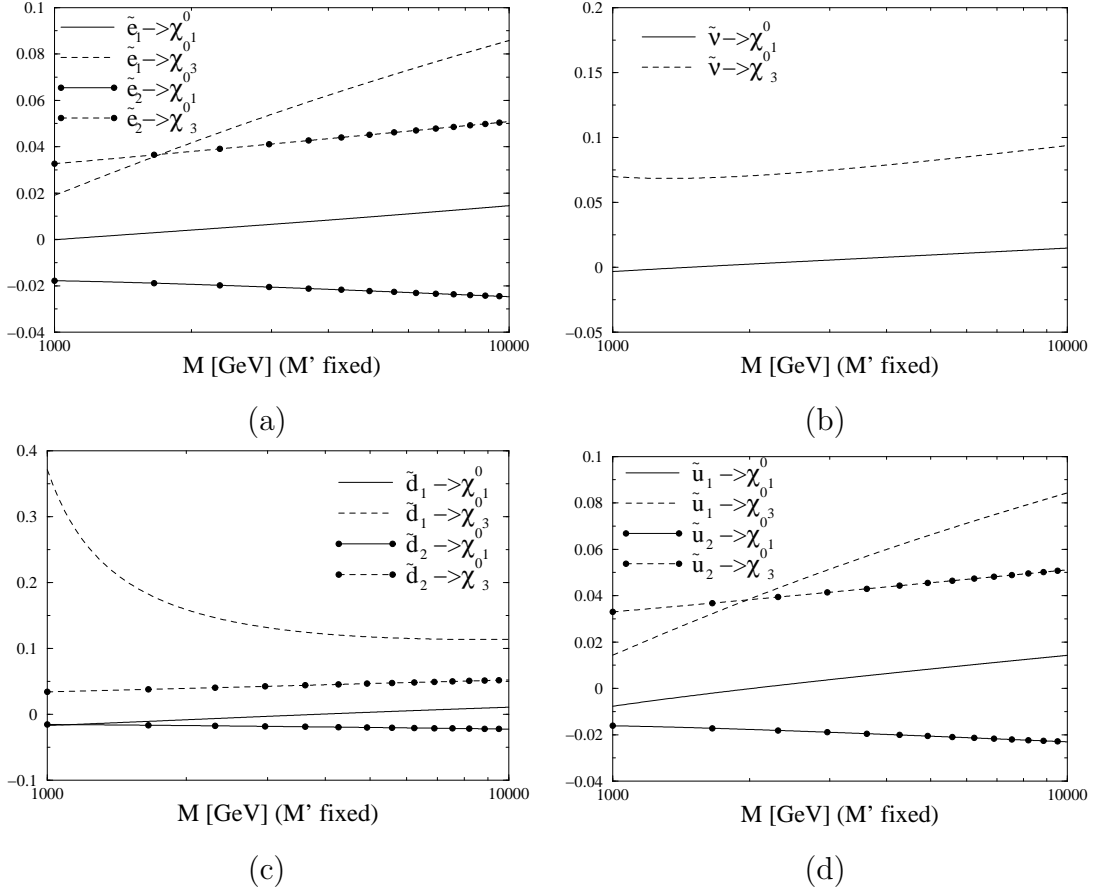
In Fig.~\ref{fig:nondecM2} we show the non-universal corrections to the
slepton and squark decays into neutralinos as a function of the
soft-SUSY-breaking $SU(2)$ gaugino mass parameter $M$ in the range
$1-10\TeV$, {keeping} $M'$ fixed to a light value $M'=120\GeV$. For
$M<1\TeV$ the corrections {show} a rich structure that will be analyzed
below.
Only the results for the first generation of sfermions are shown;
although the results for top- and bottom-squarks are slightly different,
the same conclusion follows. Shown are the corrections for each
individual decay  channel with  a branching ratio larger than $10\%$. 
For $M=1\TeV$ most of the decays have already reached the logarithmic
regime, so the figure shows mainly straight lines. The corrections can
have both signs, and range between $2\%$ and $20\%$ at $M=1\TeV$. The
slopes of the curves are small which means that, although there exist a
non-decoupling effect, it is very small and of no phenomenological
interest for $M<10 \TeV$.

A similar situation is found in the non-decoupling effect of the
$U(1)_Y$ gaugino with respect to the decays into charginos in
Fig.~\ref{fig:nondecM1}.  In this case the slopes of the different
curves are even smaller. 

The non-decoupling effects from Higgs particles can be seen in
Fig.~\ref{fig:nondecmhplus}. The figure makes clear that the corrections
grow as a $\log(\mHp)$. The effect is, of course, much larger in the
third generation squark decays, but is also visible in the $\sel$ and
$\sneut_e$ decays. 

\fignine

\subsubsection{General analysis}

We will present here the behaviour of the non-universal electroweak
corrections as a function of parameters relevant for the
chargino/neutralino sector and the sfermion sector, and \tb. In general the
corrections present a rich structure, due to the fact that every single
parameter controls different aspects of the decay under study. Let's
take for example the higgsino mass parameter $\mu$. When changing this
parameter, the neutralino and chargino masses change, and some decay
channels open or close: when this happens, threshold divergences appear
in the rest of the channels. At the same time, for $|\mu|= M$ our
renormalization prescription breaks down
{(Cf. eq.(\ref{eq:soldM2dmu}))}, and divergent corrections 
appear. Moreover the $\mu$ parameter enters the coupling of the sfermion
to Higgs bosons in two ways: directly, in the expression of the Feynman
rules, and indirectly, in the determination of the soft-SUSY-breaking
trilinear couplings~(\ref{eq:Abt}). {As a result the variation of
the corrections with respect to $\mu$ will exhibit a complicated
evolution pattern spotted with the spurious divergences associated with
the renormalization framework.}

\figten
\figeleven

\figtwelve
\figthirteen

In Figs.~\ref{fig:mulept} and~\ref{fig:muquark} we {display} 
 the
corrections for the partial decay widths into neutralinos and charginos
for  sleptons and squarks respectively. In all plots we can see the
divergences at $|\mu|=M$. We also see similar structures in several
plots. These similarities make clear the $SU(2)$ structure of the
theory, for example in the corrections to the partial decays of $\sel_1$,
$\sneut_e$, $\stau_1$, $\supq_1$ and  $\sdown_1$ into charginos. The
corresponding decays into neutralinos deviate among the different
sfermions, since the $U(1)_Y$ charge enters the game. The presence of
large Yukawa couplings also alters the general behaviour of the
corrections. The corrections to the top- and bottom-squark decays
exhibit indeed a very different structure. Differences in the $\stau$
decays would only be visible in the $\stau_2\to \sneut \cmin$  decay
channel, which has a branching ratio below $1\%$ (see table~\ref{tab:massbr}).

{Figs.~\ref{fig:Mlept} and~\ref{fig:Mquark} show the evolution with the
soft-SUSY-breaking gaugino mass parameter $M$. Again we observe a very
similar structure}
for the sleptons and the first generation squarks. For
$M<|\mu|$, and away from the divergence region, the corrections are
small (less than $5\%$) in most channels. Only the decay mode
$\stopp_2\to b\cplus$ 
has corrections around $10\%$. For $M>|\mu|$ the corrections can be
moderate, up to $5\%$ for the sleptons and first generation squarks,
and $10\%$ for the bottom-squarks. 
{Note also the divergence appearing in
the corrections to the  $\sbottom$ decays into $\cmin$ when the value of
the masses approach the phase space limit. This divergence arises from the
Coulomb singularity due to 
{soft-photon} exchange between the two final
state particles~\cite{Sommerfeld}. A consistent description of the decay
width in this mass regime needs of a proper description of  slowly
moving 
{final-state} charged particles~\cite{Sakharov}.   } 

\figfourteen

We  now turn our view to the parameters of the sfermion sector. We start
with the sfermion mixing angle. Its main effect is not in the
corrections themselves, but in the tree-level decay amplitudes. In
Fig.~\ref{fig:theta} we show the variation of the relative corrections
as a function of relevant mixing angles. For the $\stau$ decays we see
spikes of large corrections  for
the chargino channels. These spikes reflect the fact that near $\ostau=0$
($\ostau=\pm\pi/2$) the $\stau_2$ ($\stau_1$) has a tiny branching
ratio to charginos; it is basically a $\stau_R$, which (at the
tree-level) couples only to higgsino-type charginos with a small Yukawa
coupling. At one-loop, however, the $\stau_R$ does effectively couple to
charginos both through the one-loop conversion $\stau_R\to\stau_L$ and
through genuine vertex diagrams. These contributions can be of the same
size as the tree-level contributions, giving large corrections.
For the sfermions
of the first two generations these spikes are more pronounced. Since
their Yukawa couplings are negligible, the one-loop effective coupling is
larger than the tree-level one. 

The third
generation squarks have a very different behaviour. To understand
Fig.~\ref{fig:theta}b one has to bear in mind that the heaviest squark
mass varies with the squark mixing angle. As $\ost$ varies  the
$\stopp_1$ mass crosses a series of thresholds. The corrections to the
$\stopp_2$ decays on the other hand have a smooth behaviour, similar to
the $\stau$ decays but with the spikes 
softened by the larger Yukawa coupling. 

We also show in Fig.~\ref{fig:theta}c the variation of the bottom-squark
decays as a function of the $\stopp$ mixing angle $\ost$. Although the shape of
the figure might be reminiscent of what would be obtained by the use of the
finite threshold corrections to the bottom-quark Yukawa
coupling~\cite{DmbTeo,Dmb,davidDmb}, this is not the leading contribution to the
corrections. In fact, we have checked that the use of the finite
threshold corrections to the bottom-quark Yukawa coupling reproduce
quite well the shape of the corrections to the $\sbottom_2$ ($\equiv
\sbottom_R$) partial decay widths. However,  there exist  finite terms
from other contributions, changing the overall value
of the corrections. 
For the $\sbottom_1$ ($\equiv \sbottom_L$) the
finite threshold corrections fail to give an approximate description of
the full result.

\figfifteen
\figsixteen

{The evolution with the sfermion  mass parameters themselves can
  be seen in Figs.~\ref{fig:msflight}-\ref{fig:msquarkheavy}. In
Fig.~\ref{fig:msflight} we vary one of the sfermion masses, by
maintaining the splitting between $\sfr_1$, $\sfr_2$ 
as in eq.~(\ref{eq:inputpars}). In Figs.~\ref{fig:msflight}a and b we
show the non-universal {corrections to the selectron
and sneutrino partial decay widths}. Above $m_{\tilde f}\gsim 1\TeV$ the
corrections follow the Sudakov  double-log form $\delta\sim A+B
\log^2 (m_{\tilde f}^2/\mws)$~\cite{Ciafaloni:1999xg}. {{This kind of} 
electroweak corrections  
{appears} in any observable in
which the process energy is much larger than the electroweak mass
scale.}
For comparison the universal contributions are shown in
Figs.~\ref{fig:msflight}c and d  as a function of a common slepton
mass. 
{While the universal effects dominate over
 the Sudakov terms in the relevant 
region where the sfermion masses
 lie below $1\,TeV$, the opposite holds once these masses become larger.}
}

{It is
interesting to further explore the behaviour of the corrections in the
mass region below $1\TeV$ also for the squark decays.}
In Fig.~\ref{fig:msquarkheavy} we show the
corrections for top- and bottom-squark decays as a function of the
masses, in a mass range $100-600\GeV$. In this figure
the splitting between the
bottom-squarks is fixed at $5\GeV$, whereas the top-squark mass is left
free. Several thresholds are seen in the figures. The value of the
corrections behaves smoothly between the threshold points, but it is
clear that the value of the corrections depends strongly on the exact
correlation between the several MSSM masses.

\figseventeen

We come finally {to} the $\tb$ parameter. For the first two sfermion
generations $\tb$ only enters the corrections through the expressions of
the masses, since the Yukawa couplings are negligible. As a consequence
the evolution with $\tb$ is flat above $\tb\simeq4$, when the masses
{acquire} their asymptotic {values}.

For the third generation the couplings of fermions and sfermions with
Higgs bosons and higgsinos loops enter the game. 
If we stick to the input parameters in~(\ref{eq:inputpars}) the
trilinear soft-SUSY-breaking couplings $A_f$ 
{acquire} large values -- see eq.~(\ref{eq:Abt}) -- and the sfermion-Higgs
couplings become large, and even non-perturbative. 
Note however, that large
$A_f$ values would generate charge and colour breaking vacuum --
eq.~(\ref{eq:necessary}). As long as the $A_f$ are consistent with the
(necessary) condition (\ref{eq:necessary}) the corrections
{remain} 
perturbative. This is not to say that scenarios with $\tb\gsim5$ are not
possible, but that, for a given $\tb$, the input parameters $A_f$,
$m_{{\tilde f}_i}$ and $\theta_f$ are constrained so that
eq.~(\ref{eq:necessary}) is satisfied. 
{In particular},  
note that a zero
mixing angle in the sbottom sector is not possible for large $\tb$. To
{probe} the effect of $\tb$ in the corrections we must choose a
different 
parameter set that satisfies the bound~(\ref{eq:necessary}) all over
the explored $\tb$. To this end we choose the soft-SUSY-breaking masses
and trilinear couplings as input parameters, so that they reproduce the
masses and angles of~(\ref{eq:inputpars}),
{that is we
adjust the values of the parameters in the sfermion mass matrices~(\ref{eq:sbottommatrix}) as follows}
\begin{equation}
\begin{array}{l}
M_{\tilde{q}_L}=M_{\tilde{d}_R}=M_{\tilde{b}_L}=300\GeV\,\,,\,\,
M_{\tilde{u}_R}=292 \GeV\,\,,\,\,
M_{\tilde{b}_R}=299\GeV\,\,,\,\,M_{\tilde{t}_R}=274\GeV\,\,,\,\,\\
M_{\tilde{l}_L}=302\GeV\,\,,\,\,M_{\tilde{l}_R}=297\GeV\,\,,\,\,\\
A_d=A_b=A_l=600\GeV\,\,,\,\,A_u=37.5\GeV\,\,,\,\,A_t=-78\GeV\,\,,
\end{array}
\label{eq:inputSSB}
\end{equation}
{with a common value for first and second generation squarks
  ($q\equiv u,d,s,c$, $u\equiv u,c$, $d\equiv d,s$), and for all
  sleptons ($l\equiv e,\mu,\tau $).}
{The rest of the parameters are set as in
  eq.~(\ref{eq:inputpars}) ({viz.} $\mu=150 \GeV$, $M=250\GeV$,
  $\mHp=120\GeV$).} 
By choosing this parameter set, the masses and mixing angles will change
with $\tb$. In Fig.~\ref{fig:deltatb} we show the corrections to the
third generation sfermion  decays as a function of $\tb$. 
{We
see that a large value of $\tb$ may increase the absolute value of the
corrections, but they
stay below few ten percent for the allowed range $\tb<60$.}

\subsection{QCD corrections and SUSY threshold effects}
\label{sec:QCD}
{The QCD corrections to the partial decays widths of
sfermion decays into charginos and neutralinos were computed
in~\cite{Djouadi:1997wt,Kraml:1996kz}.}
We
have performed an independent computation, 
{and have
verified the analytical and numerical results of
Ref.\cite{Djouadi:1997wt}.}
{We will
need these partial results to compute the full one-loop branching ratios
of the sfermion decays.}

The QCD corrections can be very large in certain parts of the parameter
space. This applies {especially} for squarks decaying into higgsino-type
charginos or neutralinos. These large corrections can be absorbed by an
adequate resummation of the leading effects. 
{The latter} are of
two types: the running of the light-quark masses up to the scale of the
sfermion masses produces large negative corrections; the finite
SUSY threshold corrections to the bottom-quark Yukawa coupling can 
{also be} large; {they} grow with $\tb$, and its sign is
opposite to 
$\mu$ -- see the 
extensive literature~\cite{Coarasa:1996qa,DmbTeo,Dmb,davidDmb} on the subject. The authors of
Ref.~\cite{davidDmb} demonstrated that, for the case of $H^+t\bar{b}$
coupling, these corrections can be exactly resummed to all orders of
perturbation theory by {using the effective bottom quark Yukawa
  coupling according to} 
\begin{equation}
  \label{eq:mbdef}
  h_b^{eff} \equiv
  \frac{\mb^{eff}}{v_1}\equiv\frac{\mb(Q)}{v_1\,(1+\Delta_b)}\,\,, 
\end{equation}
where $\mb(Q)$ is the running quark mass, and $\Delta_b$ is the finite
threshold correction. 

By using the effective {bottom quark} mass in eq.~(\ref{eq:mbdef}) we are able to
absorb {a} large part of the corrections into the effective couplings,
yielding an 
improved  partial decay width 
\begin{equation}
  \label{eq:impQCD}
  \Gamma^{imp}\equiv\Gamma^{0}(\mb^{eff})+\left(\Gamma^{1-loop}-\Gamma^{1-expans}\right)\equiv\Gamma^{0}(\mb^{eff})\,(1+\delta^{rem})\,\,,
\end{equation}
where $\Gamma^{1-expans}$ is the one-loop expansion of
$\Gamma^{0}(\mb^{eff})$. 
{The previous
equation defines the \textit{remainder} of the one-loop contributions, namely
what is left after subtracting the one-loop part of the resummed
threshold corrections from the full one-loop result. It reads}
\begin{equation}
  \label{eq:remQCD}
  \delta^{rem}=\frac{\Gamma^{1-loop}-\Gamma^{expand}}{\Gamma^{0}(\mb^{eff})}\,\,.
\end{equation}
{The improved QCD correction factor is then given by}
\begin{equation}
  \label{eq:deltaQCD}
  \delta^{imp-QCD}= \frac{\Gamma^{imp}-\Gamma^{0}(\mb)}{\Gamma^{0}(\mb)}\,\,.
\end{equation}

\figeighteen

In Fig.~\ref{fig:QCDcorr} we show an example. 
We
have plotted the one-loop and the improved QCD corrections to the
bottom-squark partial decay widths into neutralinos as a function of
$\tb$, using the input parameters~(\ref{eq:inputSSB})
{and a 
gluino mass 
$\mg=500\GeV$}. The strong
coupling constant is evaluated at the mass scale of the decaying
particle, using the normalization $\alpha_S(\mz)=0.12$. 
 The growing {in absolute value} of
the one-loop QCD corrections is due to the increase of the
bottom-squark mass splitting {(entering $\Delta_b$)} with
$\tb$. For $\tb\gsim 25$ they already 
pass the $-100\%$ value. On the other hand $\delta^{rem}$ is well
behaved in all the $\tb$ range, {and it is seen not to be negligible at all in
some cases.}

The large negative corrections visible in
  Fig.~\ref{fig:QCDcorr} have a twofold origin. First, the running of
  the bottom quark mass provides large negative corrections due to
  \textit{standard} QCD renormalization group  effects. Second, the sign
  of the QCD  contributions to the threshold corrections is opposite to
  that of $\mu$ (since $\Delta_b\propto \mu$ in eq.~(\ref{eq:mbdef})). Therefore both {kinds} of
  contributions reinforce mutually in the {originally} chosen
  scenario $\mu>0$. {In 
  the alternative scenario ($\mu<0$) the two
  contributions partially 
  cancel each other, giving a smaller total {(negative)} correction.}
{Thus in the $\mu<0$ scenario the
SUSY-QCD effects actually prevent the decay rates from being too much
suppressed by the gluonic corrections.}

One has to be careful in the use of the effective bottom-quark
mass~(\ref{eq:mbdef}) in the case of positive corrections ($\mu<0$),
since in this case the effective Yukawa coupling grows and can become
non-perturbative for large values of $|\mu|$ and $\tb$. 
{In Fig.~\ref{fig:QCDcorr}d we show a scenario with $\mu<0$. The
input parameters are those of eq.~(\ref{eq:inputSSB}), but changing the
sign of $\mu$ and the soft-SUSY-breaking trilinear couplings
$A_f$. Although the corrections are still negative, we clearly see the
change of trend for large $\tb\gsim20$; from this point onwards the (positive)
threshold corrections are comparable to the effects of the running
bottom quark mass. At $\tb\simeq60$ the two leading contributions nearly
compensate each other, giving a total {negative} correction below
$30\%$.
{So at large $\tb$
some channels become essentially free of huge negative corrections,
whereas in the original $\mu>0$ scenario these channels were reduced 
by more than $70\%$ due to quantum effects!} }

{In Ref.~\cite{Djouadi:1997wt} the numerical results for the QCD
  corrections to the partial sbottom decay widths were much smaller than
  {those presented} in Fig.~\ref{fig:QCDcorr}. The reason is that in
  Ref.~\cite{Djouadi:1997wt} the following set of parameters was used:
  a low value of $\tb$ ($=1.6$), small splitting between the two sbottom
  masses ($=10\GeV$), and a small sbottom mixing angle
  ($\osb\simeq0$). Under these conditions the SUSY
  threshold corrections 
  to the bottom Yukawa coupling ($\Delta_b$) are
  suppressed~\cite{DmbTeo,Dmb,davidDmb}. Moreover, the choice of $\mu<0$ in
  Ref.~\cite{Djouadi:1997wt} would mean that the finite threshold
  effects {partially compensate} for the renormalization group
  running of 
  the bottom mass, {thus} decreasing the value of the corrections. 
  At moderate or large values of \tb\ 
  ($\gsim 10$), however, the rest of {the} conditions can not be
  fulfilled. 
  If the sbottom mixing angle were small ($\osb\simeq0$), then the
  soft-SUSY-breaking trilinear coupling would be large ($A_b\simeq
  \mu\tb$), eventually spoiling the vacua
  condition~(\ref{eq:necessary}). Even if one drops the
  condition~(\ref{eq:necessary}), then the weak corrections would blow
  {up} 
due to the non-perturbativity of the Higgs-squark-squark trilinear
  coupling entering the three-point function corrections in
  Fig.~\ref{diags:3point}. We conclude, therefore, that sticking to the
  necessary condition~(\ref{eq:necessary}) provides a natural suppression
  of the Higgs-squark-squark trilinear coupling, {while}
  maintaining the 
  perturbativity of the weak corrections over the whole $\tb$ range. At
  the same time, it means that {for} 
 $\tb\gsim10$ the two sbottom masses
  have {significant} splitting, and $\osb\neq0$, providing large
  corrections due to the finite threshold effects. }

{We remark that
whereas the finite threshold corrections ($\Delta_b$) can be regarded as
\textit{non-decoupling} effects in the study of the Higgs sector of the
MSSM, this is not so for the sfermion decays under consideration (even
though we have kept the \textit{non-decoupling} denomination also in our
case). Indeed, the non-decoupling property}
  applies when \textit{all} the SUSY parameters ($\mu$, $A_b$, $\msba$,
  $\mg$) are simultaneously scaled.
{However, being the sbottom quark mass itself the process
energy}, the scaling of the SUSY parameters entails a
  simultaneous scaling of the process energy, 
{and therefore the $\Delta_b$ corrections cannot be considered
here as genuine non-decoupling quantum effects.} 
{Moreover,  in an scenario of large $\mu$ ($\mu \gg \msba$) the light
chargino
is basically a gaugino, and  so its coupling to quarks and squarks is
essentially a gauge coupling,
not a Yukawa coupling.  As a result the sbottom decay into light
charginos/neutralinos 
is not sensitive to $\Delta_b$ in this limit.}

\subsection{Branching ratios}

{At the end of the
day we analyze the higher-order effects on the branching ratios. Notice
that in practice only the branching ratios are directly accessible from
the measurements of the various cross-sections at the colliders. Therefore the
$BR(\tilde{f} \to f' \chi)$ are the true observables in this kind of
analysis, and we will compute them from the corresponding corrections to
the partial decay widths that we have considered in the previous
sections. In particular we have to include the QCD and the threshold
effects. In the following analysis we also 
take into  account the (generally small) effects of the shift
{of} the masses in the {phase-space}
factor. Specifically we show  the 
following quantities:}
\begin{itemize}
\item Tree-level branching ratio ($BR^0$);
\item QCD-correction to the branching ratio as discussed in
  section~\ref{sec:QCD}, including the QCD corrections to 
  the $\stopp_1$ {mass}:
\begin{equation}
\Delta BR^{QCD}(\sfr \to \chi)=
\frac{\Gamma^{imp}(\sfr\to\chi)}{\sum_{\chi^\pm_i,\chi^0_\alpha} \Gamma^{imp}(\sfr\to\chi)}-BR^0(\sfr\to\chi)\ \,\,,
\end{equation}
{$\Gamma^{imp}$ being defined in~(\ref{eq:impQCD})};
\item {Total} correction to the branching ratios (QCD and EW), including the QCD
  and EW corrections to the $\stopp_1$ {mass}, and the EW
  corrections to 
  the neutralino masses 
\begin{equation}
\Delta
BR^{total}(\sfr\to\chi)=\frac{\Gamma^{imp}(\sfr\to\chi)+\Gamma^0(\sfr\to\chi)
\delta( \sfr_a\to f'\chi)}{
\sum_{\chi^\pm_i,\chi^0_\alpha} \left( \Gamma^{imp}(\sfr\to\chi)+\Gamma^0(\sfr\to\chi)
\delta( \sfr_a\to f'\chi)\right)}-BR^0(\sfr\to\chi)\,\,,
\end{equation}
{with $\delta(\sfr_a\to f'\chi)$ as defined in~(\ref{eq:totalcn}).}
\end{itemize}

The corrections to the branching ratios are usually smaller
than those to the partial decay widths. The {limiting} case
{being}  when only one
decay channel exists, and has obviously no corrections to
the branching ratio. 

Only results for the bottom- and top-squarks will be shown. The
corrections to the slepton and first- and second-generation squarks are
tiny, for two reasons: usually a single decay channel is dominant
{and}  the
corrections to the partial decay widths are small. In addition the
corrections will be shown only on those portions of the parameter space
in which the only possible decay channels are the fermionic ones. The
bosonic decay channels $\sq_a\to\sq'_b (V,H)$ can be dominant when they
are open, and the quantum corrections to these channels must be taken
into account to compute the corrections to the branching ratios~\cite{Bartl:1998xk}. 

\fignineteen

\figtwenty

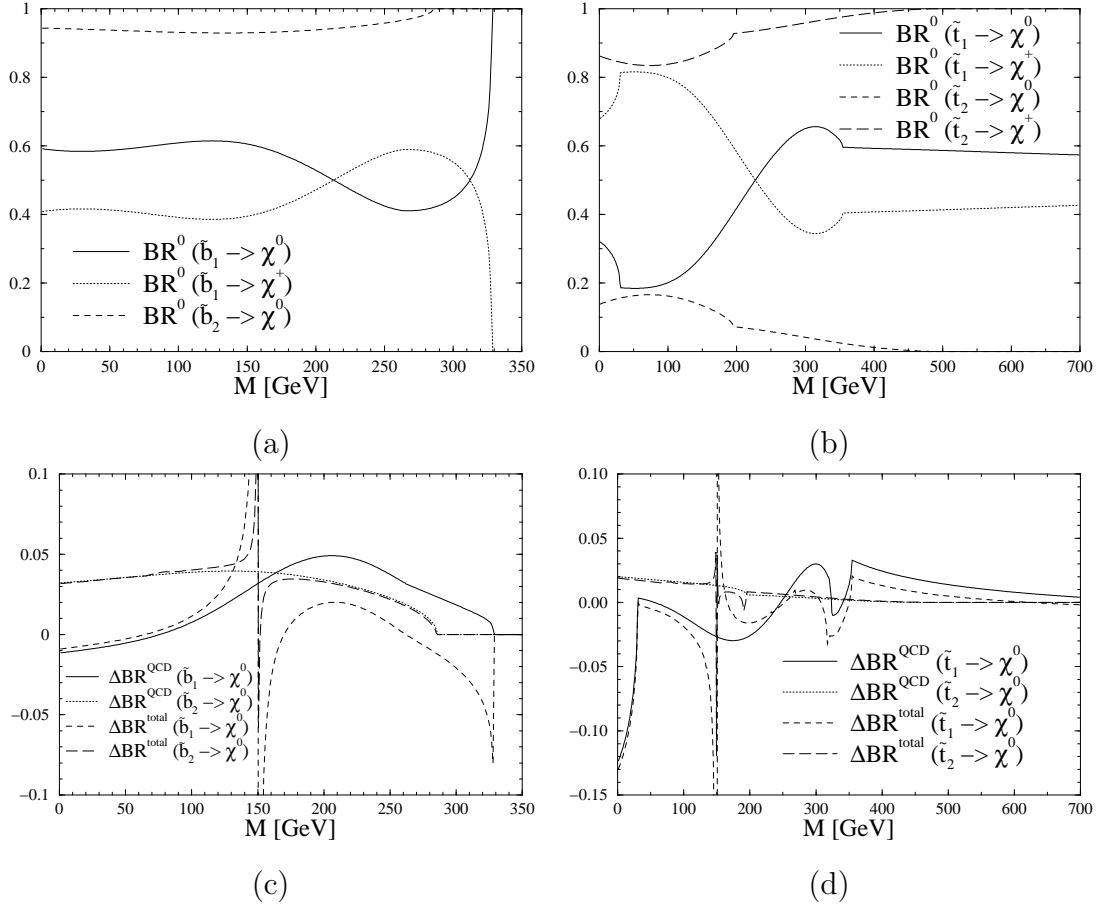
In Fig.~\ref{fig:brM} we show the tree-level branching ratios of the
bottom- and top-squarks as a function of the $SU(2)_L$
soft-SUSY-breaking gaugino mass parameter $M$, as well as the
absolute corrections to the branching ratios of squarks decaying into
neutralinos\footnote{Since only two decay channels are open, the
  corresponding absolute corrections to the 
  chargino branching ratio have the same absolute value and opposite
  sign.}.
Although the QCD corrections are usually the largest ones, the EW and
QCD corrections can be of the same order in certain scenarios. For
example in Fig.~\ref{fig:brM}c the QCD corrected $BR(\sbottom_1\to
b\neut)$ is $\sim5\%$ larger than the tree-level one for $M\sim200-250\GeV$,
but the EW corrections compensate most of this correction, and the final
correction is less than $1\%$.

The effects of the mass-shifts are in general very
small, except near the threshold regions, where a given channel is
permitted according to the tree-level masses prediction, but it is
closed when one uses the one-loop prediction for the masses. This is the
case in Fig.~\ref{fig:brM}d. The decay channel $\stopp_1\to b\neut_3$
is open up to $M\simeq 354\GeV$ according to the tree-level prediction
for the heaviest top-squark mass. {However} the negative corrections to
$\msto$ --table~\ref{tab:sup2masses}-- 
{enforce this channel to get closed for}
lighter $M$ values, concretely, at $M\simeq 324\GeV$ including only the
QCD corrections and at $M\simeq317\GeV$ including the full corrections. 

In Fig.~\ref{fig:brmsb2} we show the branching ratios and its
corrections as a function of the lightest bottom-squark mass
({keeping}
the splitting between the two bottom-squarks at
$\msbo=\msbt+5\GeV$). The lines representing the corrections end 
at the points 
where the bosonic decay channels open. 
We see again that the EW corrections to the $\sbottom_1$ branching ratio
can be of the same order {as} 
the QCD ones. For $\msbt\lsim 320\GeV$
the  $3-4\%$ QCD contributions are almost compensated by the EW
ones. For the $\sbottom_2$, the EW corrections represent a small shift to
the QCD-corrected branching ratios. For the top-squark the QCD corrections
can be larger ($\gsim 10\%$), and the branching ratio  also suffers an
important shift 
from the EW sector.

\figtwentyone

The effects of the mixing angles are shown in Fig.~\ref{fig:brosq}. 
The corrections to the bottom-squark branching ratio show a simple
structure. The corrections from the weak sector are comparable to that
of the QCD sector {for practically any value
 of the mixing angle}. The $\stopp$
branching ratios and corrections are shown in Figs.~\ref{fig:brosq}b and
d. The effect of the EW corrections is clearly visible above the
QCD-corrected branching ratios. 

\figtwentytwo

We come finally to analyze the $\tb$ dependence, which is shown in
Fig.~\ref{fig:brotb}. In this case we will use 
again the soft-SUSY-breaking mass parameters in eq.~(\ref{eq:inputSSB})
to compute the physical masses and mixing angles in the sfermion
sector. The main effect of $\tb$ on the tree-level branching ratios is
due to the 
change in the sfermion masses themselves, providing the opening of the
bosonic channels. Again, the weak corrections to the bottom-squark
branching ratios are seen to be a tiny addition to the QCD induced
ones. Opposite to that, the weak corrections to the top-squark decay
branching ratios are larger, {especially} in the lightest top-squark
channels. In this case they can be comparable to the QCD induced
corrections. 

\section{Conclusions}
\label{sec:conclu}

We have presented a consistent and complete {one-loop} on-shell
renormalization scheme for the 
sfermion and the chargino-neutralino sectors of the MSSM. This scheme is
suitable for the computation of one-loop electroweak corrections to
observables relevant to the next generation of colliders.

We have applied this scheme to compute the {full}
 electroweak
corrections to the partial  decay widths of sfermions into charginos and
neutralinos. 

\tablethree

 As a summary of the results we show in Table~\ref{tab:brcorr}
  the corrected branching ratios for all possible {individual}
  sfermion decays for the {(low \tb)} input 
  parameter set~(\ref{eq:inputpars}). For squarks we show both: the
  branching {ratio} including only QCD effects, and the fully corrected
  branching ratio. One can compare these results with
{the former tree-level results in}
  Table~\ref{tab:massbr} to {assess} the importance of the
  corrections in 
  this particular parameter set\footnote{We have not searched for
  an \textit{optimized} parameter set to enhance or decrease the value
  of the corrections, but just used \textit{typical} values for the
  input parameters. The reader is warned that in other \textit{typical}
  scenarios (e.g. large \tb) the corrections may look much different, as
  we have shown in previous sections.}. A close comparison shows that,
  in fact, the EW corrections can have an effect as large as  the QCD
  corrections alone. We can compare, for example,  the leading branching
  ratio of the  lightest   up-squark ($\supq_2\to u\neut_1$: $BR^0=58\%$,
  $BR^{QCD}=57.9\%$, $BR^{total}=56.1\%$) or the lightest sbottom
  ($\sbottom_2\to b\neut_1$: $BR^0=50.2\%$, $BR^{QCD}=54.1\%$,
  $BR^{total}=52.8\%$).

 The corrections show a rich {and complicated} structure when the MSSM
parameters are varied. 
 {Nevertheless, we {have been} able to provide physical
   explanations 
   of our 
 results.}

{Indeed}, we have identified  sources of non-decoupling effects in
{these} radiative corrections. 
{In
 contradistinction to the Standard Model case,} 
none of the MSSM particles decouples from these
corrections\footnote{This applies to supermultiplets
  in which one of the components is light, e.g.\ a SM particle. A
  supermultiplet in which all of the components are heavy will show
  decoupling properties.} {-- the
non-decoupling effects being 
logarithmic in the heavy 
masses. As a consequence  all} particles of the MSSM must be
taken into account in any computation involving {loop diagrams
  with external} 
fermion-sfermion-chargino/neu\-tra\-li\-no {couplings}.
In some cases, however, the term multiplying this logarithm is
small. 

{Furthermore,} we have identified a class of universal corrections to the
fermion-sfermion-chargino/neu\-tra\-li\-no couplings, that can be treated as
\textit{effective coupling matrices} for the chargino/neu\-tra\-li\-no
sector. These universal corrections consist of the fermion-sfermion
contributions to the self-energies of the gauge bosons, Higgs bosons,
charginos and neutralinos. The \textit{effective coupling matrices} absorb
the non-decoupling effects of sfermions. Explicit analytic expressions
for these corrections have been given in a simple example, {but
  the general analysis has been 
  performed numerically.}
These corrections can be large
($\sim 5-10\%$ for sfermion masses around $1\TeV$) and grow
logarithmically with the sfermion masses.  {A physical
  explanation of this effect using renormalization group arguments has
  been given.}

The {bulk of the} non-universal corrections grow as a logarithm squared of the
particle decay mass, due to the Sudakov-type double logarithms of the
electroweak corrections.

For sfermion masses around $300\GeV$, relevant for a $800\GeV$ $e^+e^-$ linear
collider such as TESLA, the non-universal electroweak corrections to the
slepton and first- and second-generation squark partial decay widths are
small. For top- and bottom-squarks the corrections are larger due to the
large Yukawa couplings. {The \textit{threshold}-like corrections
to the quark Yukawa couplings are important at large \tb\  for 3rd
generation sfermions, and they must be {resummed} to obtain
meaningful results.} 

The corrections {remain} always {in the perturbative
  regime} as long as 
the  soft-SUSY-breaking trilinear couplings are not too
large. This condition is, however, granted if the vacuum {does
  not break the} charge and/or colour {symmetry}. 
{In
 particular, we point out that this vacuum condition cannot be preserved
 at large $\tb$  by degenerating the sfermion masses and/or assuming
 vanishing sfermion mixing angles. Therefore, since $\mu=0$ is
 phenomenologically ruled out, 
 one can
 compensate the $\mu\tb$ term in eq.~(\ref{eq:Abt}) by an appropriate
 choice of the other 
 parameters. For instance,  at $\tan\beta=30$
 the vacuum condition can be preserved with the following set of masses and
 mixing angles:
 $m_{\sbottom}=(334,268) \GeV$, $\osb\simeq -0.743$
 $m_{\stopp}=(356,308) \GeV$ and $\ost\simeq -0.576$.}

The corrections can be significantly larger for individual decay
channels that have small branching ratios. Therefore large corrections
are washed out in the total decay widths $\Gamma(\sfr \to f\neut)$ and
$\Gamma(\sfr \to f'\chi^{\pm})$ -- eq.~(\ref{eq:totalcn}).

{We have combined the QCD corrections with the electroweak effects for
the top- and bottom-squark partial decay widths, and have evaluated the
full one-loop branching ratios of these supersymmetric particles in the
case that the only open decay channels are the chargino/neutralino
ones. In performing the computation of the corrections to the decay
rates and branching ratios we have also taken into account the
corrections to the squark and neutralino masses themselves. An specially
interesting case appears when these mass shifts provide the opening
(closing) of channels that would be closed (open) according to the naive
tree-level prediction. Since the overall corrections can be very large
for higgsino-type charginos/neutralinos, we have made use of the
resummed expressions for the two leading quantum contributions to the
bottom-quark Yukawa coupling, namely the running quark mass and the
finite threshold supersymmetric effects.}

{The upshot of our analysis should be emphasized: the EW corrections can
be of the same order of magnitude as the QCD effects, and therefore a
consistent treatment of the sfermion decays beyond leading order in the
MSSM demands to include the EW quantum contributions on the same footing
as the QCD ones.}

\section*{Acknowledgments}
The calculations have been done using the QCM cluster of the 
DFG Forschergruppe ``Quantenfeldtheorie, Computeralgebra und
Monte-Carlo Simulation''.
We are thankful to T. Hahn for
his help regarding the Computer Algebra system.
J.G. is thankful to D. St{\"o}kinger, A. Vicini, G.
Rodrigo, M. Melles and M. Spira for useful discussions.
This collaboration is part of the network ``Physics at Colliders'' of the
European Union under contract HPRN-CT-2000-00149.
The work of J.G. is supported by the 
European Union under contract No. HPMF-CT-1999-00150. 
The work of J.S.
has been supported in part by MECYT and FEDER under project FPA2001-3598.

\providecommand{\href}[2]{#2}\begingroup\raggedright
\endgroup

\end{document}